\documentclass[10pt,journal,compsoc]{IEEEtran}
\usepackage{caption}
\usepackage{xurl}
\usepackage{lipsum}
\usepackage{algorithmic}
\usepackage[linesnumbered,lined,boxed,commentsnumbered]{algorithm2e}
\usepackage{multicol}
\usepackage[utf8]{inputenc}
\usepackage{amssymb}
\usepackage{subfigure}
\usepackage{color}
\usepackage{url}
\usepackage{multirow}
\usepackage{epsfig}
\usepackage{epstopdf}
\usepackage{graphicx}
\usepackage{subfig}
\usepackage{amssymb}
\usepackage[figuresright]{rotating}
\usepackage{float}

\ifCLASSOPTIONcompsoc
  \usepackage[nocompress]{cite}
\else
  \usepackage{cite}
\fi
\ifCLASSINFOpdf
\else
\fi
\begin{document}
%
\title{Privacy in targeted advertising: A survey}
%
%
%
%

\author{Imdad Ullah,~\IEEEmembership{Member, IEEE,} Roksana Boreli, and Salil S. Kanhere,~\IEEEmembership{Senior Member, IEEE}
\IEEEcompsocitemizethanks{\IEEEcompsocthanksitem I. Ullah is with the College of Computer Engineering and Sciences, Prince Sattam bin Abdulaziz University, Al-Kharj 11942, Saudi Arabia.\protect\\
E-mail: i.ullah@psau.edu.sa
\IEEEcompsocthanksitem R. Boreli was with CSIRO Data61 Sydney, Australia.\protect\\
E-mail: roksana@tmppbiz.com
\IEEEcompsocthanksitem S. S. Kanhere is with UNSW Sydney, Australia.\protect\\
E-mail: salil.kanhere@unsw.edu.au}
}

\markboth{IEEE Communications Surveys \& Tutorials}%
{Shell \MakeLowercase{\textit{et al.}}: Bare Demo of IEEEtran.cls for Computer Society Journals}
%



\IEEEtitleabstractindextext{%
\begin{abstract}

Targeted advertising has transformed the marketing landscape for a wide variety of businesses, by creating new opportunities for advertisers to reach prospective customers by delivering personalised ads, using an infrastructure of a number of intermediary entities and technologies. The advertising and analytics companies collect, aggregate, process and trade a vast amount of user's personal data, which has prompted serious privacy concerns among both individuals and organisations. This article presents a detailed survey of the associated privacy risks and proposed solutions in a mobile environment. We outline details of the information flow between the advertising platform and ad/analytics networks, the profiling process, advertising sources and criteria, the measurement analysis of targeted advertising based on user's interests and profiling context and the ads delivery process, for both \textit{in-app} and \textit{in-browser} targeted ads; we also include an overview of data sharing and tracking technologies. We discuss challenges in preserving user privacy that include threats related to private information extraction and exchange among various advertising entities, privacy threats from third-party tracking, re-identification of private information and associated privacy risks. Subsequently, we present various techniques for preserving user privacy and a comprehensive analysis of the proposals based on such techniques; we compare the proposals based on the underlying architectures, privacy mechanisms and deployment scenarios. Finally, we discuss the potential research challenges and open research issues.

\end{abstract}

}

\maketitle

\IEEEdisplaynontitleabstractindextext

%
\IEEEpeerreviewmaketitle

\section{Introduction} \label{section-introduction}



Online advertising has become a prevalent marketing tool, commanding the majority of spending and taking over from the traditional broadcast advertising in newspapers, or television and radio. This is primarily due to the ability of online ad platforms to tailor or personalise ads, and thereby target specific customer segments. Targeted advertising is based on Big data analytics, where user's personal information is collected and processed to enable segmenting users into groups based on interests, location, or personal attributes like age, gender, etc., with a varying size of the selected customer segment, down to the level of an individual.

The most significant platform from which personal data are collected and subsequently used for targeted ads is a mobile device, including mobile phones or tablets, due to it's widespread and almost continuous use by a huge audience of potential ad recipients. A recent report \cite{grnm} lists that 69\% of user's digital media time is actually spent on mobile phones only and consequently recommends tailoring targeted ads for mobile devices. Although the mobile users are still utilising browsers to access various online sites, applications (\textit{apps}) are increasingly replacing the generic browser functionality. Currently, millions of mobile \textit{apps} can be downloaded via various \textit{app} marketplaces like the Google Play Store and the Apple App Store; it is projected that there will be more than 250 billion mobile \textit{app} downloads by the end of 2021 \cite{appsDownloads}.

Most mobile \textit{apps} contain at least one ad library (including analytics\footnote{Analytics is the systematic computational analysis of data or statistics for deeper understanding of consumer requirements. E.g. Google Analytics \url{https://analytics.google.com}, Flurry Analytics \url{https://www.flurry.com/analytics/}.} libraries) \cite{grace2012unsafe} that enables targeted (or behavioural) mobile advertising to a wide range of audiences.  The information about users and their online behaviour is collected through the ad library API calls \cite{book2013case}, including information inference based on monitoring ads displayed during browsing sessions \cite{Chaabane:NDSS:2012, castelluccia2012betrayed}. The Advertising and Analytics (A\&A) companies like Google Analytics and Flurry use this framework and are competing to increase their revenue by providing ad libraries that the \textit{apps} developers use to serve ads. In the process of data monetisation, the avertising/analytics companies aggressively look for all possible ways to gather personal data from users, including purchasing users' personal data from third parties.

The collection and use of personal data poses serious threats to privacy of users \cite{ullah2017enabling, ullah2020protecting, tchen2014, mamais2019privacy, liu2016privacy, wang2020aiming}, when websites or \textit{apps} indicating sensitive information are used as the basis for profiling, e.g., a gaming \textit{app} showing a gambling problem. Privacy concerns have been increasingly recognised by policy makers, with the introduction of anti-tracking laws, gradually making the use of some third-party tracking techniques used for interest-based targeting obsolete. E.g. Google has announced the Chrome's 'Cookie Apocalypse', planning to phase out support for third-party cookies by 2022\footnote{\url{https://www.adviso.ca/en/blog/tech-en/cookie-apocalypse/}}. Subsequently, instead of relying on third- party data, the A\&A companies are increasingly using first-party data and shifting towards maintaining their own Data Management Platforms (DMPs) and Demand-Side Platforms (DSPs)\footnote{DMP is a unified and centralised technology platform used for collecting, organising, and activating large sets of data from disparate sources. DSP allows for advertisers to buy impressions across a number of different publisher sites, all targeted to specific users and based on key online behaviors and identifiers. See \url{https://www.lotame.com/dmp-vs-dsp/} for detailed discussion over DMP and DSP.} to brand their own data and measure performance in a `cookie-less' world. In a stronger push towards increased user's privacy control over collection and use of their data, Apple\footnote{\url{https://junction.cj.com/article/button-weighs-in-what-does-apples-idfa-opt-in-overhaul-mean-for-affiliate}}  has recently introduced the Identification for Advertisers (IDFA) opt-in overhaul in iOS 14.5, which will have significant impact on targeted ads and mobile ad/data attribution. This has created a very public feud with one of the largest social networks (and private data collection companies), Facebook \cite{cnet}, highlighting two different business approaches in regards to privacy and user targeting.

Overall, regardless of the technological and policy changes, protecting users' personal data while having effective targeting is important to both the advertising networks and mobile users. Mobile users do want to view relevant (interest-based) ads, provided that their information is not exposed to the outside world including the advertising companies. Advertising networks can only be effective if they deliver the most relevant ads to users, to achieve better view/click through rates, while protecting the interactions between mobile users, advertisers and publishers/ad networks.

In this paper, we survey the threats and solutions related to privacy in mobile targeted advertising. We first present a survey of the existing literature on privacy risks, resulting from the information flow between the A\&A companies, temporal tracking of users regarding both their activities and the outcomes of targeting them with personalised ads. We then describe, for both \textit{in-app} (note that we interchangeably use `mobile' and `\textit{in-app}') and \textit{in-browser} targeted ads: the user profiling process, data collection and tracking mechanism, the ad delivery process and the process of ad characterisation. We outline the privacy threats posed by the A\&A companies as a result of targeting; in particular, (to prove the privacy leakage) we demonstrate, using experimental evaluation, how private information is extracted and exchanged among various entities in an advertising system including third-party tracking and highlight the associated privacy risks. Subsequently, we provide an overview of privacy preserving techniques applicable to online advertising, including differential privacy, anonymisation, proxy-based solutions, k-anonymity i.e. generalisation and suppression, obfuscation, and crypto-based techniques such as Private Information Retrieval (PIR) and blockchain-based techniques. We also survey the proposed privacy preserving advertising systems and provide a comparative analysis of the proposals, based on the underlying architectures, the privacy techniques used and the deployment scenarios. Finally, we discuss the research challenges and open research issues.

This article is organised as follows. In Section \ref{section1-mobileads}, we introduce the mobile advertising ecosystem, its operation for ad delivery process, profiling process and characterisation of \textit{in-app} and \textit{in-browser} ads. Section \ref{section-ads-operations} discusses the technical and in-depth understanding of ad network operations for targeted ads. Section \ref{section-challenges} presents privacy threats and information leakage in online advertising systems. Section \ref{solutions} presents a detailed comparative analysis of various privacy-preserving advertising systems. Various open research issues are outlined in Section \ref{open-issues}. We conclude in Section \ref{conclusion}.

\section{The Mobile Advertising Network}\label{section1-mobileads}
The ad network ecosystem involves different entities which comprise of the advertisers, ad agencies and brokers, ad networks delivering ads, \textit{analytics} companies, publishers and the end customers to whom ads are delivered \cite{leontiadis2012don}. For the case of large publishers, the ads may be served both by the publishers and the advertisers \cite{vallina2012breaking}, consequently, the ad ecosystem includes a number of interactions between different parties.

\subsection{The advertising ecosystem} 
\begin{figure}[h]
\begin{center}
\includegraphics[scale=0.5]{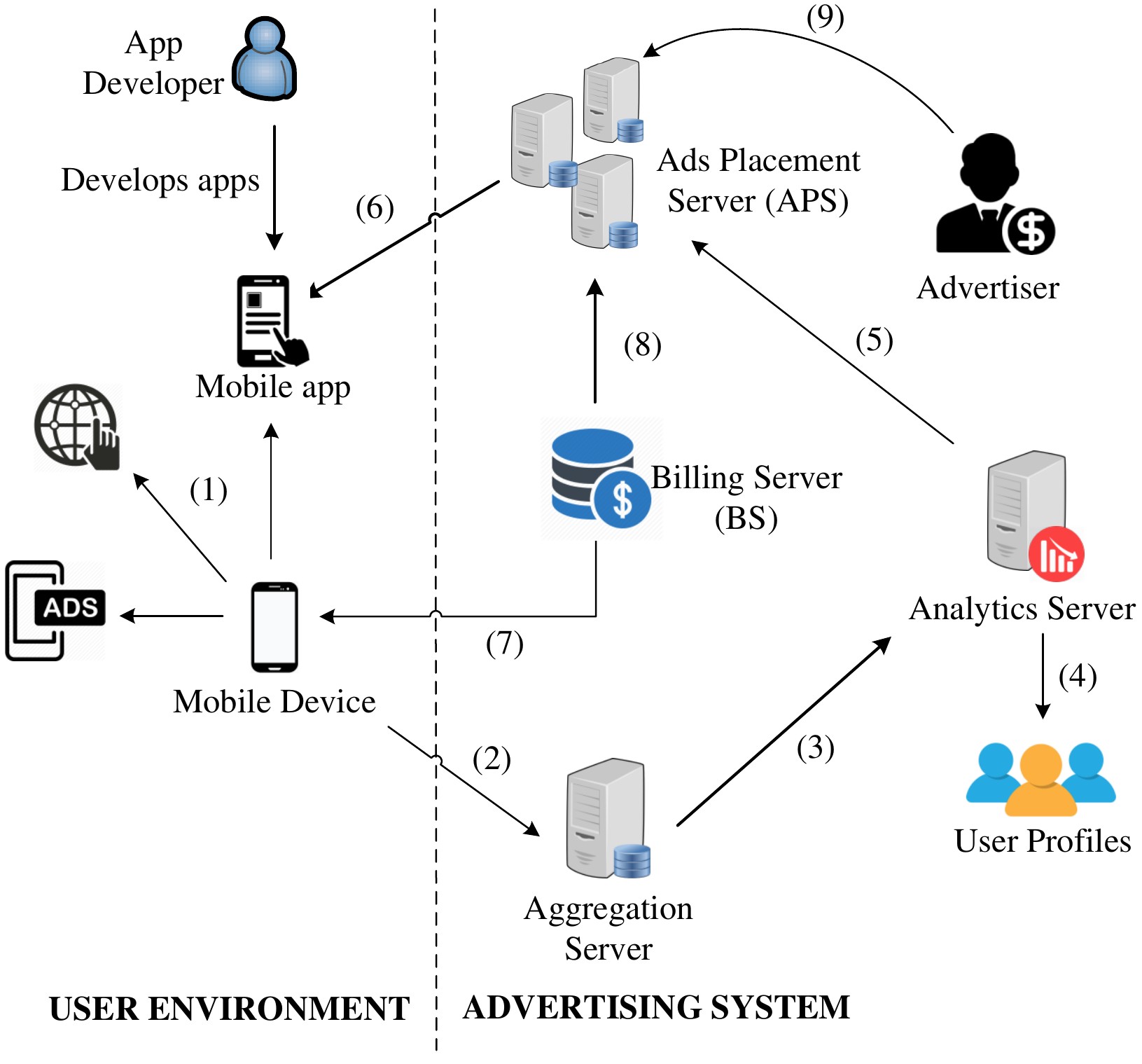}
\caption{The mobile advertising ecosystem, including the information flow among different parties. Following functionalities: (1) Data collection and tracking, (2) Send tracking data to \texttt{Aggregation} server, (3) Forward usage info to \texttt{Analytics} server, (4) User profiling, (5) Send profiling info to \texttt{APS}, (6) Deliver targeted/generic ads, (7) Billing for \textit{apps} developer, (8) Billing for Ad System, (9) Advertiser who wishes to advertise with Ad system.}
\label{fig:attack2}
\end{center}
\end{figure}

A typical mobile ad ecosystem (both for \textit{in-app} and \textit{in-browser} ads) and the information flow among different parties is presented in Figure \ref{fig:attack2}. A user has a number of \textit{apps} installed on their mobile device, that are utilised with specific frequency. As demonstrated in \cite{han2012study}, most mobile \textit{apps} include \textit{analytics} Software Development Kit (SDK) and as such both report their activity and send ad requests to the \textit{analytics} and ad network. This network comprises the \texttt{Aggregation} server, \textit{analytics} server, \texttt{Billing} server, and the \texttt{Ads Placement Server} (APS). Collected data, that relates to usage of mobile \textit{apps} and the success of displayed ads, is used by the ads \textit{\textit{analytics}} server to develop user profiles (associated with specific mobile devices and corresponding users). A user profile comprises a number of \textit{interests}, that indicates the use of related \textit{apps}, e.g. sports, business, etc., constructed by e.g., Google \textbf{Ad}vertising network for \textbf{Mob}ile (AdMob)\footnote{Google AdMob profile is accessible through the \textit{Google Settings} system \textit{app} on Android devices, accessible through \texttt{Google Settings} $\to$ \texttt{Ads} $\to$ \texttt{Ads by Google} $\to$  \texttt{Ads Settings}.} and Flurry \cite{flurry} (note that the latter is only visible to \textit{app} developers). \textit{Targeted} ads are served to mobile users according to their individual profiles. We note that other i.e., \textit{generic} ads are also delivered \cite{ullah2014characterising}. The \texttt{Billing} server includes the functionality related to monetising \textit{Ad impressions} (i.e. ads displayed to the user in specific \textit{apps}) and \textit{Ad clicks} (user action on selected ads); further discussion over ads \textit{billing} is given in Section \ref{billing-ads}.


\subsection{User profiling}
Advertising systems rely on user \textit{profiling} and \textit{tracking} to tailor ads to users with specific interests and to increase their advertising revenue. Following, we present the user \textit{profiling} process, in particular, how the user profile is \textit{established}, various criteria, and how it \textit{evolves} over time.

\subsubsection{Profile establishment}
\begin{figure}[h]
\begin{center}
\includegraphics[scale=0.4]{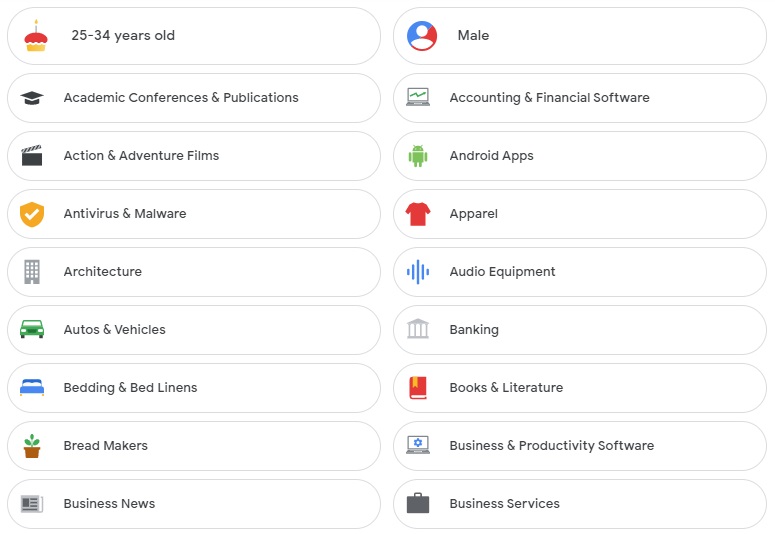}
\caption{An (anonymous) example user profile estimated by Google as a results of \textit{Web \& App} activity.}
\label{example-profile}
\end{center}
\end{figure}

The advertising companies, e.g., Google, profile users based on the information they add to their Google account, data collected from other advertisers that partner with Google, and its estimation of user's interests based on mobile \textit{apps} and websites that agree to show Google ads. An example profile estimated by Google with various demographics (e.g. gender, age-ranks) and profiling interests (e.g. Autos \& Vehicles) is shown in Figure \ref{example-profile}. It is assumed that there is a \textit{mapping} of the \textit{Apps profile} $K_a$ (the \textit{apps} installed on a user mobile device) to an \textit{Interests profile} $I_g$ (such an example set of interests is shown in Figure \ref{example-profile}) defined by advertising (e.g. Google) and \textit{analytics} companies i.e. $K_a \to I_g$. This information is used by the \textit{analytics} companies to individually characterise user's interests across the advertising ecosystem.

This \textit{mapping} includes the conversion of the \textit{apps} categories $\Phi _j$ (where $j=1,..., \tau $ and $\tau $ is the number of different categories in a marketplace) to interest categories $\Psi_l$ ($l=1,..., \epsilon $.  $\epsilon $ is the number of interest categories defined by the \textit{analytics} company). This \textit{mapping} converts an \textit{app} ${a_{i,j}} \in {S_a}$ to interests set $S_g^{i,j}$ after a specific level of activity ${t_{est}}$. The ${t_{est}}$ is the  \textit{establishment threshold} i.e. time an \textit{app} should be used in order to establish profile's interests. The result of this \textit{mapping} is a set of interests, called \textit{Interests profile} $I_g$. Google profile interests\footnote{Google profile interests are listed in \url{https://adssettings.google.com/authenticated?hl=en}, displayed under the 'How your ads are personalized'. Note that Google services can also be verified on Google Dashboard \url{https://myaccount.google.com/dashboard?hl=en}.} are grouped, hierarchically, under vaiours interests categories, with specific interests.

In addition, the ads \textit{targeting} is based on demographics so as to reach a specific set of potential customers that are likely to be within a specific age range, gender etc., Google\footnote{Demographic Targeting \url{https://support.google.com/google-ads/answer/2580383?hl=en}} presents a detailed set of various \textit{demographic targeting} options for ads display, search campaigns etc. The demographics $D$ are usually grouped into different categories, with specific options such as age-ranges, e.g. `18-24', `25-34', `35-44', `45-54', `55-64', `65 or more', and gender e.g., `Male', `Female', `Rather not say', and other options e.g. household income, parental status, location etc. The \textit{profiling} is a result of interactions of user device with the AdMob SDK \cite{ullah2020protecting} that communicates with Google \textit{analytics} for deriving user profiles. A complete set of `\textit{Web \& App} activities' can be found under `My Google Activity'\footnote{\url{https://myactivity.google.com/myactivity?otzr=1}}, which helps Google make services more useful, such as, helping rediscover the things already searched for, read, and watched.

\begin{figure*}[h]
\begin{center}
\includegraphics[scale=0.7]{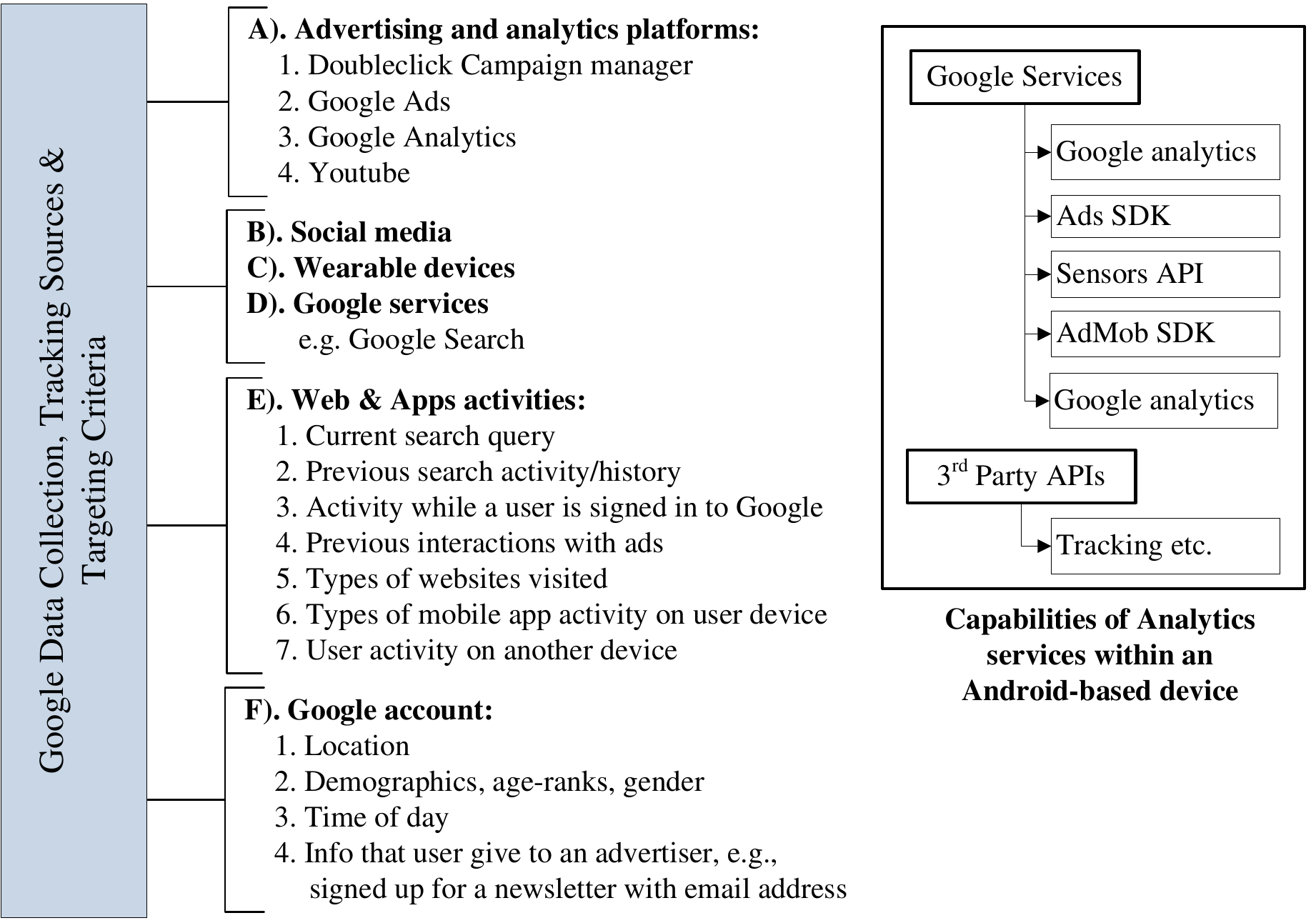}
\caption{Google's data collection and \textit{tracking} sources for \textit{targeting} users with personalised ads (left) and tracking capabilities of analytics libraries enabled within mobile devices (right).}
\label{data-extraction}
\end{center}
\end{figure*}

Figure \ref{data-extraction} shows, a specific example of Google, various sources/platforms that Google use to collect data and \textit{target} users with personalised ads. These include a wide range of different sources enabled with various tools, e.g., the `\textit{Web \& Apps} activities' are extracted with the help of Andoird/iOS SDKs, their interactions with \textit{analytics} servers within Google network, cookies, \textit{conversion tracking}\footnote{\url{https://support.google.com/google-ads/answer/6308}}, web searches, user's interactions with received ads etc. Similarly, Google's connected home devices and services\footnote{Google's Connected Home Devices and Services: \url{https://support.google.com/googlenest/answer/9327662?p=connected-devices&visit_id=637357664880642401-2675773861&rd=1}} rely on data collected using cameras, microphones and other sensors to provide helpful features and services\footnote{Sensors in Google Nest devices: \url{https://support.google.com/googlenest/answer/9330256?hl=en}}. Google Takeout\footnote{\url{https://takeout.google.com/}} can be used to export a copy of contents (up to several GBs of data) in user's Google Account for backup or use it with a service outside of Google. Furthermore, this includes the data from a range Google products personalised for specific users that a user use, such as, email conversations (including `Spam' and `Trash' mails), contacts, calendar, browsing \& location history, and photos.

%
%

\subsubsection{Profile evolution} The profile is updated, and hence the ads \textit{targeting}, each time variations in the users' behavior are observed; such as for a mobile user using \textit{apps} that would map to interests other than the existing set of interests. Let a user uses a new set of \textit{apps} $S{'_a}$, which has no overlap with the existing set of \textit{apps} $S_a$ that has created $I_g$ i.e., $S'_a \subset {\mathcal{A}}\setminus {S_a}$, $\mathcal{A}$ is the set of \textit{apps} in an \textit{app} market. The newly added set of \textit{apps} $S{'_a}$ is converted to interests with ${t_{evo}}$ as \textit{evolution threshold} i.e. the time required to evolve profile's interests. Hence, the final \textit{Interests profile}, $I_g^f$, after the \textit{profile evolution} process, is the combination of older interests derived during the profile \textit{establishment} $I_g$ and during when the profile \textit{evolves} $I{'_g}$.

\subsubsection{Profile development process} 
\begin{figure}[h]
\begin{center}
\includegraphics[scale=0.4]{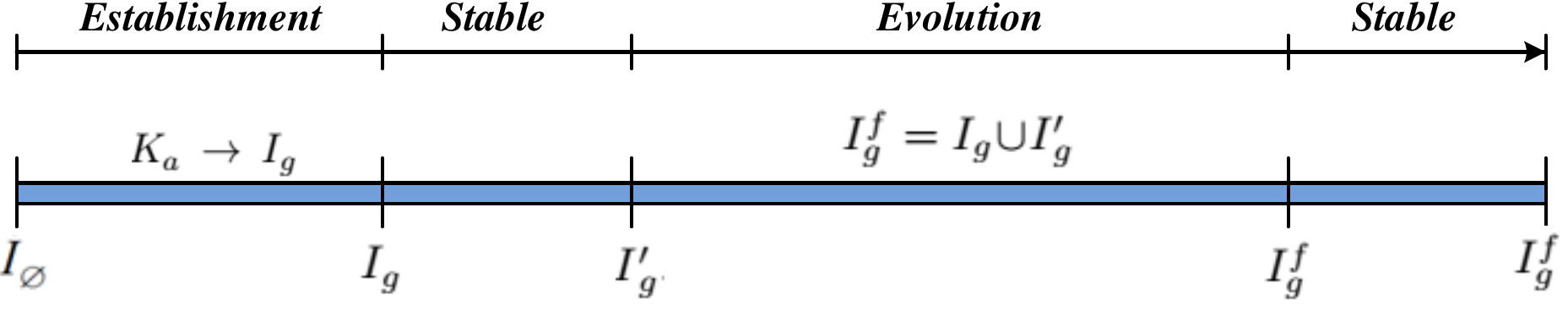}
\caption{Profile \textit{establishment}  \& \textit{evolution} processes. ${I_\emptyset }$ is the empty profile before \textit{apps} utilisation. During the \textit{stable} states, the \textit{Interest profiles} $I_g$ or $I_g^f$ remains the same and further activities of the same \textit{apps} have no effect over the user profiles.}
\label{figure-profileEvoEstb}
\end{center}
\end{figure}

In order for the \textit{Apps profile} to \textit{establish} an \textit{Interests profile}, a minimum level of activity of the installed \textit{apps} is required. Furthermore, in order to generate one or more interests, an \textit{app} needs to have the AdMob SDK. We verified this by testing a total of 1200 \textit{apps} selected from a subset of 12 categories, for a duration of 8 days, among which 1143 \textit{apps} resulted the \textit{Interest profiles} on all test phones indicating ``Unknown'' interests. We also note that the \textit{Apps profile} deterministically derives an \textit{Interests profile} i.e., a specific \textit{app} constantly derives identical set of interests after certain level of activity. We further note that the level of activity of installed \textit{apps} be within a minimum of 24hours period (using our extensive experimentations; we note that this much time is required by Google analytics in order to determine ones' interests), with a minimum of, from our experimentations,  $24/n$ hours of activity of $n$ \textit{apps}. For a sophisticated \textit{profiling}, a user might want to install and use a good number of \textit{apps} that would represent one's interests. After the 24hours period, the profile becomes \textit{stable} and further activity of the same \textit{apps} does not result in any further changes. The mapping of \textit{Apps profile} to \textit{Interests profile} during the \textit{establishment} and during the \textit{evolution} process along with their corresponding \textit{stable} states are shown in Figure \ref{figure-profileEvoEstb}.

Similarly, during the profile \textit{evolution} process, the \textit{Interests profile} starts changing by adding new interests; once \textit{apps} other than the existing set of \textit{apps} $S_a$ are utilised. However, instead of 24hours of period of evolving a profile, we observe that the \textit{evolution} process adds additional interests in the following 72hours of period, after which the aggregated profile i.e. $I_g^f$ becomes \textit{Stable}. In order to verify the stability of the aggregated profile, we run these \textit{apps} on 4th day; henceforth we observe no further changes. The mapping of \textit{Apps profile} to \textit{Interests profile} during the \textit{establishment} and during the \textit{evolution} process along with their corresponding \textit{Stable} states are shown in Figure \ref{figure-profileEvoEstb}.

\subsection{Targeted advertising}
The mobile targeted advertising is a crucial factor in increasing revenue (a prediction shows the mobile ad market to grow to \$408.58 billion in 2026 \cite{mobstastistics}) in a mobile \textit{app} ecosystem that provides free services to the smartphone users. This is mainly due to users spend significantly more time on mobile \textit{apps} than on the traditional web. Hence, it is important (note that \textit{targeted} advertising is not only unique to the mobile ads but has also been used for \textit{in-browser} to deliver ads based on user's interests. The characterisation of \textit{targeted} advertising, on the user's side, is the in-depth analysis of the ad-delivery process so as to determine what information the mobile \textit{apps} send to the ad network and how effectively they utilise this information for ads \textit{targeting}. Furthermore, the characterisation of mobile targeted ads would expose the ad-delivery process and the ad networks can use the resultant analysis to enhance/redesign the ad delivery process, which helps in better view/click through rates.

It is crucial for the \textit{targeted} advertising to understand as what information do \textit{apps} (both free and paid mobile \textit{apps} of various categories) send to the ad networks, in particular, how effectively this information is used to \textit{target} users with interest-based ads? whether the ad networks differentiate among different types of users using \textit{apps} from the same or different \textit{apps} categories (i.e. according to \textit{Apps profile})? how much the ad networks differentiate mobile users with different profiles (i.e. according to \textit{Interests profile})? the effect over user \textit{profiling} with the passage of time and with the use of \textit{apps} from diverse \textit{apps} categories (i.e. during profile \textit{evolution} process)? the distribution of ads among users with different profiles? and the frequency of unique ads along with their ads serving distributions?

\subsection{Ads selection algorithms}
The accurate measurement of the \textit{targeted} advertising is systematically related to the ad selection algorithm and is highly sensitive since it combines several fields of mathematics, statistics, analytics, and optimisation etc. Some of the ad selection algorithms show ad selection based on the user data pattern \cite{ng2002intelligent} and the program event analysis \cite{thawani2004event}, however, the \textit{contextual} and \textit{targeted} advertising is treated in different way as they are related to the psyche of the users. Consequently, it has been observed that the activity of users and their demographics highly influences the ad selection along with the user clicks around an ad \cite{yan2009much, jaworska2008behavioural}. As an example, a young female that is frequently browsing websites or using mobile \textit{apps} related to the category of \textit{entertainment}, would be more interested in receiving ads related to \textit{entertainment} such as movies, musical instruments etc., consequently, it increases the \textit{click-through rates}. Another work \cite{shin2019targeted} builds a game-theoretic model for ad systems competing through \textit{targeted} advertising and shows how it effects the consumers' search behavior and purchasing decisions when there are multiple firms in the market. We note that the researchers utilise different ad selection and \textit{targeting} algorithms based on machine learning and data mining techniques.

\subsection{Ad billing}\label{billing-ads}
Billing is an important part of business models devised by any advertising system that is based on billing their customers for fine grained use of ad systems and their resources e.g. the advertisers set the payment settings and payment methods for monetising \textit{ad impressions} and \textit{clicks}. A number of studies show potential privacy threats posed by billing \cite{danezis2011differentially, balasch2010pretp, henry2011practical} i.e. a privacy-invasive architecture consists of service provides collecting usage information (such as particular interests of ads being shown and clicked) in order to apply appropriate tariff.n Hence, among the important aims of private billing is to eliminate the leakage of private information and to minimise the cost of privacy across the \textit{billing} period.

\begin{figure}[h]
\begin{center}
\includegraphics[scale=0.99]{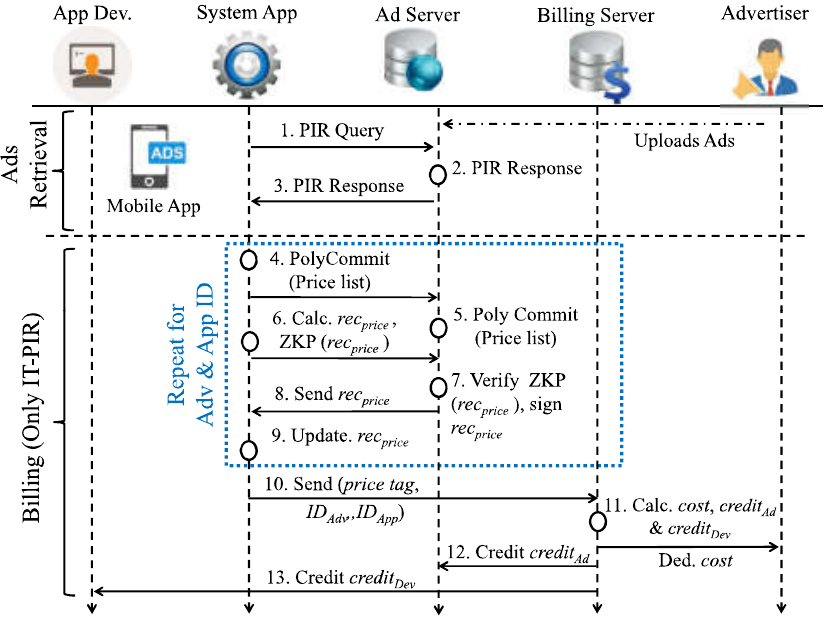}
\caption{The work flow for Ads retrieval and billing for \textit{ad presentations} and \textit{ad clicks} \cite{ullah2017enabling}.}
\label{private-billing-pir}
\end{center}
\end{figure}

An example implementation of our private \textit{billing} for ads, based on \textit{ZKP} and \textit{Polynomial commitment} (see detailed discussion over these techniques in Appendix \ref{private-billing}), is presented in \cite{ullah2017enabling}, also shown in Figure \ref{private-billing-pir}. In this proposal, we presume that the following information is available to the \textit{client} (software e.g. the AdMob SDK that is integrated in mobile \textit{apps} for requesting ads and tracking user's activity) for all ads in the database: the \textit{Ad} index $m$, \textit{Ad} category \({\Phi _i}\), \textit{price tags} \(C_T^{prs}\) and \(C_T^{clk}\) respectively for \textit{ad presentations} and \textit{ad clicks}, and and the \textit{Advertiser ID} \(I{D_{Adv}}\). This private \textit{billing} mechanism consists of two parts: the work flow for retrieving ads (Step 1--3) and private \textit{billing} (Step 4--13). In Step 2, the \texttt{Ad server} calculates the PIR response and sends it back to the \textit{client}, following, the \textit{client} decodes the PIR response (step 3) and forwards the retrieved ads to the mobile \textit{app}.

Once the \textit{ads presentation} (or \textit{ad click}) process finishes then it undergoes the \textit{billing} process. The \textit{client} calculates the \textit{receipt} locally, consisting of various components that are used to verify the following: (\textbf{a}) price tier for ad presented or ad clicks; (\textbf{b}) the \(I{D_{Adv}}\) (used for price deduction from advertiser, as shown in Step 11 of Figure \ref{private-billing-pir}); and (\textbf{c}) the application ID (helpful for price credit to \textit{App Developer} i.e. Step 13). This \textit{billing} mechanism is based on PS-PIR \cite{henry2011practical}, proposed for \textit{e-commerce}. We note that this \textit{billing} mechanism is only applicable to single ad requests with no impact on privacy.

\sloppy As opposed to above implementation, we suggested another proposal \cite{ullah2020privacy} for \textit{ad presentations} and \textit{clicks} with the use of mining Cryptocurrency (e.g. Bitcoin). The major aim for this proposal was for preserving user privacy, secure payment and for compatibility with the underlying \textit{AdBlock} proposal \cite{ullah2020privacy} for mobile advertising system over Blockchain. Following notations are used in this proposal: price tags \(C_{prs}^{A{d_{ID}}}\) and \(C_{clk}^{A{d_{ID}}}\) for ad \textit{presentation} and \textit{click}; various \textit{wallets} i.e. \textit{App Developer}'s \(walle{t_{I{D_{APP}}}}\), \textit{Advertiser}'s \(walle{t_{A{D_{ID}}}}\), \texttt{Billing} server's \(walle{t_{BS}}\); \textit{public-private key} ($PK+/-$) and (Bitcoin) addresses, i.e. \(Ad{d_{I{D_{APP}}}},Ad{d_{A{D_{ID}}}},Ad{d_{BS}}\). It works as follows: The advertiser buys advertising \textit{airtime}, it signs the message with the amount of Cryptocurrency with her \textit{private key} ($PK-$), adds \textit{Billing} server's address, requesting a transaction. Following, this request is bind with other transactions and broadcasted over the network for \textit{mining}. Once the transaction completes, the \texttt{Billing} server receives its portion of Cryptocurrency in her \textit{wallet}. In addition, the \texttt{Miner} initiates \textit{billing} transaction for ads \textit{presentations} or \textit{clicks} respectively by encoding the \(C_{prs}^{A{d_{ID}}}\) and \(C_{clk}^{A{d_{ID}}}\) price tags; this amount is then shared with \(walle{t_{I{D_{APP}}}}\) and \(walle{t_{A{D_{ID}}}}\) \textit{wallets}.

\section{Operations of Advertising System}\label{section-ads-operations}
Following, we discuss the technical aspects of the advertising systems e.g. the ad delivery process, ads traffic extraction and its characterisation, which eventually helps in understanding privacy issues in \textit{targeted} advertising.

\subsection{Ad delivery process}
\begin{figure}[h]
\begin{center}
\includegraphics[scale=0.37]{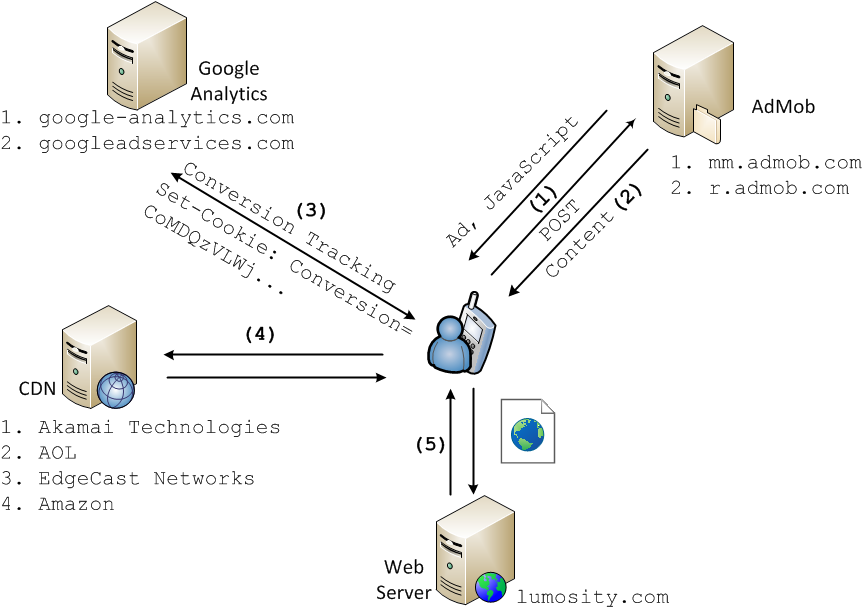}
\caption{AdMob Ad Presentation Workflow \cite{ullah2014characterising}.}
\label{figure-admob}
\end{center}
\end{figure}

We identify the workflow of a mobile \textit{app} requesting a Google AdMob ad and the triggered actions resulting from e.g. a user click (we note that other advertising networks, such as Flurry, use different approaches/messages to request ads and to report ad clicks). Figure \ref{figure-admob} describes some of the domains used by AdMob (Google ad servers and AdMob are shown separately for clarity, although both are acquired by Google). As shown, an ad is downloaded after the \texttt{POST} method is sent by mobile phone (Step 2) containing phone version, model, \textit{app} running on phone etc. The ad contains the landing page (web address of an ad-URL) and \texttt{JavaScript} code that is executed where some of the static objects are downloaded (such as a \texttt{PNG}, (Step 3)). Two actions are performed after clicking an ad: a \textit{Conversion} cookie\footnote{\textit{Conversion tracking} is specifically used by Google that is an action a customer takes on website that has value to the business, such as a purchase, a sign-up, or a view of a key page \cite{conversationTracking}.} is set inside phone (Step 4) and the web server associated with the ad is contacted. The landing page may contain other list of servers (mainly residing in Content Delivery Networks) where some of the static objects are downloaded and a complete \texttt{HTML} page is shown to the user (Step 5). The mobile \textit{apps} developers agree on integrating ads in mobile \textit{apps} and the ads are served according to various rules set by the ad networks, such as to fill up their advertising space, and/or obtaining \textit{profiling} information for \textit{targeting}. Additionally, the ads refreshment intervals, mechanisms used to deliver ads (push/pull techniques), the strategy adopted after ad is being clicked, and click-through rates etc. are also defined by the ad networks.

In consequence, the ad networks are complex systems being highly diverse with several participants and adopting various mechanisms to deliver ads. Thus, in order to correctly identify and categorise ads and to server appropriate ads, it needs to investigate various ad delivery mechanisms and also cope with such diversity. This evaluation process needs identifying and collecting various ads delivery mechanisms through inspecting collected traffic traces captured from several \textit{apps} executions, as shown in Figure \ref{figure-admob}. In addition, it needs to emphasis on ads distribution mechanisms used by ad networks from the \textit{apps}' perspective or users' interests to devise the behaviour of ads pool served from ad networks and how they map to individual user's interest profiles. Since the advertising system is a closed system, this process needs to indirectly evaluate the influence of different factors on ad delivery mechanisms, which is even more complicated in Real Time Bidding (RTB) scenarios and associated privacy risks.


\subsection{Understanding ad network's operation}
The advertising networks provide an SDK for integrating ads inside the mobile \textit{apps} while securing the low level implementation details. The ad networks provide regulation for embedding ads into the mobile \textit{apps}, the ad delivery mechanism, the amount of times an ad is displayed on the user screen and how often an ad is presented to the user. The common type of ad is the flyer, which is shown to the user either at the top or at the bottom of device's screen, or sometimes the entire screen is captured for the whole duration of the ad presentation. These flyers are composed of text, images and the \texttt{JavaScript} codes.

The ad presentation workflow of Google AdMob is shown in Figure \ref{fig:attack2} that shows the flow of information for an ad request by an \textit{app} to AdMob along with the action triggered after the user clicks that particular ad. This figure shows the \texttt{HTTP} requests and the servers (i.e. Content Delivery Network (CDN) or ad servers) used by AdMob. Furthermore, several entities/services and a number of \texttt{HTTP} requests to interact with the ad servers and user agent can be observed in this figure.

\subsection{Ad traffic analysis}

\subsubsection{Extracting ad traffic}
Recall that the mobile ad network involves different entities to interact during the ad presentation and after an ad is being clicked to download the actual contents of the ad, as observed in Figures \ref{fig:attack2} and \ref{figure-admob}. Specifically, these entities are the products, the ad agencies attempting ad campaigns for the products, ad networks delivering ads, the publishers developing and publishing mobile \textit{apps}, and the end customer to whom ads are delivered \cite{leontiadis2012don}. It is likely, when it comes to large publishers, that both the publishers and advertisers may have their own ad servers, in which case, some publishers may configure to put certain ads pool on the advertisers' side and, at the same time, maintain their own ad servers \cite{vallina2012breaking}. The publishers, this way, can increase their revenue by means of providing redundant ad sources as if one ad network fails to deliver ads then they can try another ad network to continue providing services. In similar way, an end user may experience to be passed over several ad networks from publishers to the advertisers to access ads.

\subsubsection{Ads traffic identification}
The advertising system itself and its functionality are very diverse and complex to understand its operation \cite{ullah2017enabling, ullah2020joint}, hence in order to categorise the ad traffic, it needs to be able to incorporate such diversity. This can be performed by first capturing the traces from the \textit{apps} that execute and download the ad traffic and then investigating the traffic characteristics. Characterising and inspecting the ad traffic can give information about the approaches used by multiple publishers, the various mechanisms used to deliver ads by the publishers, the use of different ad servers, and the ad networks themselves \cite{ullah2020privacy}. Similarly, it helps identify any \textit{analytics} traffic used by the ad networks to \textit{target} with relevant ads. Analysis of the traffic traces enables to parse and classify them as traffic related to \textbf{i)} ad networks, \textbf{ii)} the actual web traffic related to ad, \textbf{iii)} traffic related to CDN, \textbf{iv)} \textit{analytics} traffic, \textbf{v)} tracking traffic, \textbf{vi)} ad auctions in RTB, and \textbf{viii} statistical information about \textit{apps} usage or developer's statistics, and \textbf{ix)} traffic exchange during and after an ad click. As a consequence, a major challenge is to be able to derive comprehensive set of mechanisms to study the behaviours of ad delivery, classify the connection flows related to different ad networks, detecting any other possible traffic, and to classify them in various categories of ads.

\subsubsection{Mobile vs. in-browser ads traffic analysis}
We note that there are several differences in separately collecting and analysing the mobile and \textit{in-browser} user's ad/data traffic for the ad delivery mechanism in order to \textit{target} users. Analysing the mobile ad traffic requires to be able to derive comprehensive set of rules to study the ad delivery behaviours (since several ad networks adopt their own formats for serving ads, as mentioned above), catalogue connection flows, and to classify ads categorisation. Furthermore, the ad delivery mechanisms are not publicly available, hence, analysing mobile targeted ads would be dealing with an inadequate information problem. Although \textit{in-browser} ad delivery mechanism can be customised\footnote{E.g. by modifying Google ads preferences: \url{https://adssettings.google.com/authenticated?hl=en}} to receive ads which are tailored to a specific profiling interests \cite{guha2009privad, toubiana2010adnostic}. 

For the \textit{in-app} ads delivery \cite{rafieian2020targeting, ullah2014profileguard, ullah2020protecting, ullah2017enabling, gu2018secure}, an ad network may use different information to infer users' interests, in particular, the installed applications together with the device identifier to profile users and to personalise ads pool to be delivered. Similarly, for \textit{in-browser} ads, user \textit{profiling} is performed by \textit{analytics} companies \cite{trzcinskianalyse} through different information such as browsing history, web searches etc., that is carried out using configured cookies and consequently \textit{target} users with personalised ads. However, in \textit{in-app} ad context, this information might be missing, or altogether not permitted by the OS, as the notion of user permissions may easily prevent the access to data out of the \textit{apps} environment.

\subsection{Characterisation of \textit{in-app} advertisements}
There is a limited research available on characterising the \textit{in-app} (mobile) targeted ads. Prior research works have demonstrated the large extent to which \textit{apps} are collecting user's personal information \cite{leontiadis2012don}, the potential implications of receiving ads to user's privacy \cite{castelluccia2012betrayed} and the increased utilisation of mobile device resources \cite{khan2013cameo, vallina2012breaking}. In our previous study \cite{ullah2014characterising} (and in \cite{nath2015madscope}), we observe that various information sent to the ad networks and the level of ads \textit{targeting} are based on communicated information, similarly, we \cite{tchen2014} investigate the installed \textit{apps} for leaking targeted user data. To combat these issues, a number of privacy preserving \cite{guha2009privad, toubiana2010adnostic, haddadi2010mobiad} and resource efficient mobile advertising systems \cite{khan2013cameo, vallina2012breaking} have been proposed. Works on the characterisation of mobile ads have primarily focused on measuring the efficiency of \textit{targeted} advertising \cite{yan2009much}, to examine whether the \textit{targeted} advertising based on the users' behaviour leads to improvements in the \textit{click-through rates}. However, thus far there have been limited insights about the extent to which \textit{targeting} is effective in mobile advertising that will ultimately determine the magnitude of various issues such as bandwidth usage, including the loss of privacy.

We note that existing approaches on characterising \textit{targeted} advertisements for \textit{in-browser} \cite{balebako2012measuring, wills2012understanding, toubiana2010adnostic, castelluccia2012betrayed, guha2009privad, yan2009much, goldfarb2011online, farahat2012effective, evans2009online, barford2014adscape} cannot be directly applied to the evaluation of \textit{in-app} ads due to the following reasons: \emph{\textbf{First}}, the \textit{in-app} targeting may be based on a number of factors that go beyond what is used for \textit{in-browser} ads, including mobile \textit{apps} installed on the device, the way they are utilised (e.g. heavy gamers may receive specific ads). \emph{\textbf{Second}}, the classification of ads requires unifying of mobile market place(s) and traditional online environments, as the ads may relate both to merchant websites and to other \textit{apps} that may be purchased and downloaded to the mobile devices. \emph{\textbf{Third}}, the methodology for collecting information about \textit{in-app} ads is different than for the \textit{in-browser} ads, since the ad delivery process for \textit{in-app} ads changes with every other ad network. \emph{\textbf{Finally}}, \textit{apps} come with pre-defined \textit{apps} permissions to use certain resources, hence, allowing \textit{apps} to filter part of the information to be provided to the ad network.


\begin{figure*}[t]
\begin{center}
\includegraphics[scale=0.6]{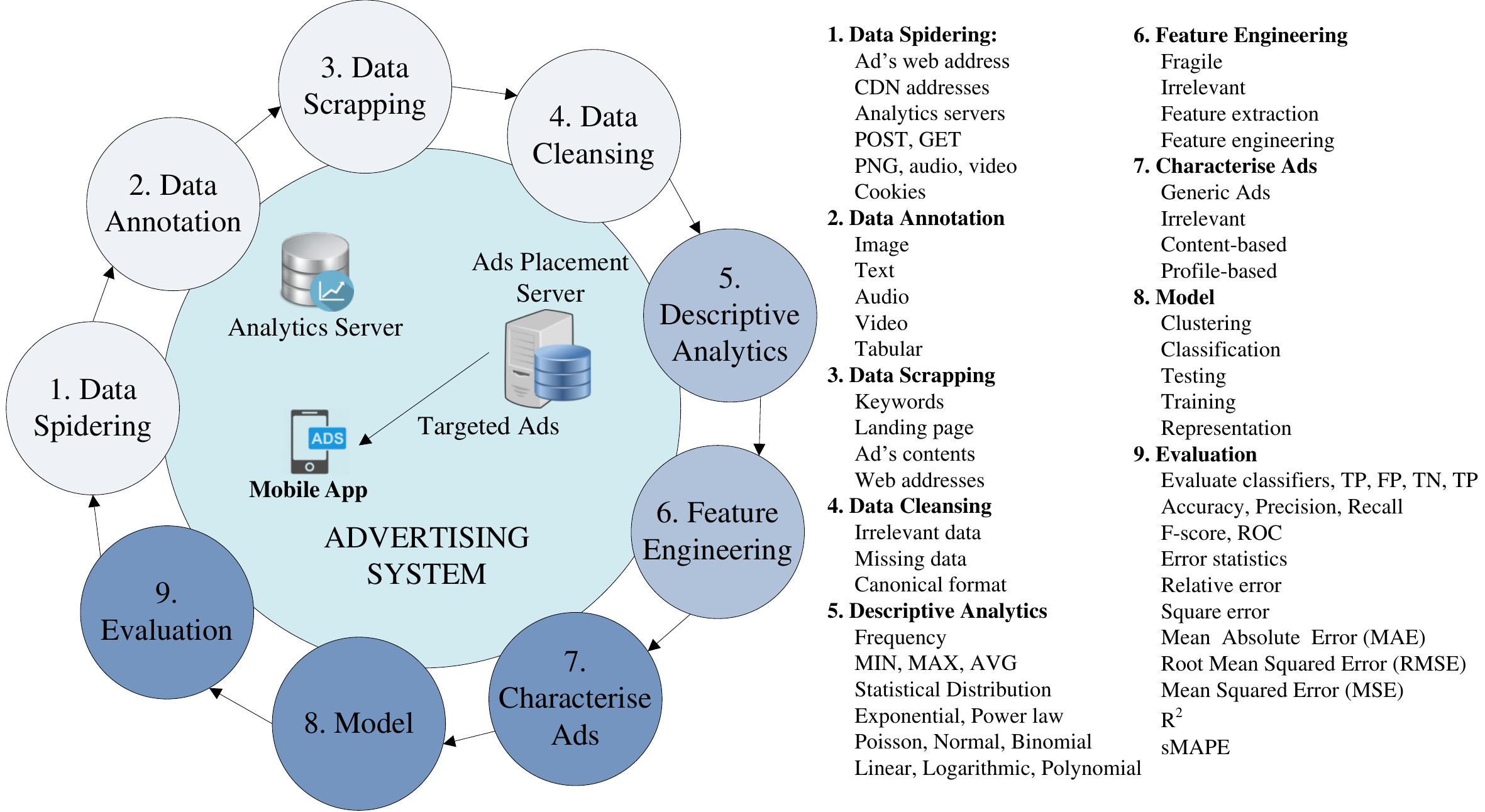}
\caption{The process of ads characterisation for both \textit{in-app} and \textit{in-browser} targeted ads. Various steps for preparing data for ads characterisation are given from `1' through `6', ads characterisation is done via '7', various models van be applied given in `8', finally, various evaluation metrics are given in `9'.}
\label{ads-chracterization}
\end{center}
\end{figure*}

Figure \ref{ads-chracterization} shows the lifecycle of characterising the ads traffic within the advertising system, both for \textit{in-app} and \textit{in-browser} \textit{targeted} ads; various data scrapping elements and statistical measures are also shown on the right side of this figure.

Following we discuss few works on the characterisation of \textit{in-app} and \textit{in-browser} targeted ads.

\subsubsection{In-app (mobile) ads}
Few studies characterise various features of \textit{in-app} ad traffic with the focus on \textit{targeted} advertising. The MAdScope \cite{nath2015madscope} and \cite{ullah2014characterising} collects data from a number of \textit{apps}, probes the ad network to characterise its \textit{targeting} mechanism and reports the \textit{targeted} advertising using profiles of specific interests and preferences. The authors in \cite{khan2013cameo} analyse the ads harvested from 100+ nodes deployed at different geographic locations and 20 Android-based phones and calculated the feasibility of caching and pre-fetching of ads. The authors in \cite{vallina2012breaking} characterise the mobile ad traffic from numerous dimensions, such as, the overall traffic, the traffic frequency, and the traffic implications in terms of, using well-known techniques of pre-fetching and caching, energy and network signalling overhead caused by the system. This analysis is based on the data collected from a major European mobile carrier with more than three million subscribers. The \cite{mohan2013prefetching} shows similar results based on the traces collected from more than 1,700 iPhone and Windows Phone users.

The authors in \cite{xu2011identifying} show that \textit{apps} from the same category share similar data patterns, such as geographic coverage, access time, set of users etc., and follow unique temporal patterns e.g. entertainment \textit{apps} are used more frequently during the night time. The \cite{lee2011study} performs a comparative study of the data traffic generated by smartphones and traditional internet in a campus network. Another work \cite{zhang2012expensive}, studies the cost overhead in terms of the traffic generated by smartphones that is classified into two types of overheads i.e. the portion of the traffic related to the advertisements and the \textit{analytics} traffic i.e. traffic transmitted to the third-party servers for the purpose of collecting data that can be used to analyse users' behaviour etc. Several other works, \cite{pathak2012energy, pathak2011fine, qian2011profiling}, study \textit{profiling} the energy consumed by smartphone \textit{apps}.

\subsubsection{In-browser ads}
There are a number of works on characterising \textit{in-browser} ads with the focus on issues associated with the user privacy \cite{evans2009online, goldfarb2011online}. In \cite{castelluccia2012betrayed}, the authors present classifications of different trackers such as cross-site, in-site, cookie sharing, social media trackers, and demonstrate the dominance of \textit{tracking} for leaking user's privacy, by reverse engineering user's profiles. They further propose a browser extension that helps to protect user's privacy. Prior research works show the extent to which consumers are effectively tracked by third parties and across multiple \textit{apps} \cite{razaghpanah2018apps}, mobile devices leaking \textit{Personally Identifiable Information} (\textit{PII}) \cite{elsabagh2020firmscope, ren2016recon} and \textit{apps} accessing  user's  private and sensitive information through well defined APIs \cite{verderame2020reliability}. Another study \cite{lecuyer2014xray} reveals by using differential correlation technique in order to identify various \textit{tracking} information used for \textit{targeted} ads. Similarly, \cite{gandhi2006badvertisements} investigates the ad fraud that generates spurious revenue affecting the ad agencies. In addition, other studies, such as \cite{guha2010challenges} describes challenges in measuring online ad systems and \cite{barford2014adscape} provides a general understanding of characteristics and changing aspects of advertising and \textit{targeting} mechanisms used by various entities in an ad ecosystem.

\section{Privacy in Mobile Advertising: Challenges} \label{section-challenges}


Privacy can be defined as ``the ability of an individual or group to seclude themselves or information about themselves, and thereby express themselves selectively\footnote{\url{https://en.wikipedia.org/wiki/Privacy}}''. In addition, the Personally Identifiable Information (PII) is the ``the  information  that  can  be  used  to  distinguish  or  trace  an  individual's identity\footnote{\url{https://www.osec.doc.gov/opog/privacy/PII_BII.html}}'', which if compromised or disclosed without authorisation, may result in harm, embarrassment, inconvenience, or unfairness to an individual. Recall that the \textit{profiling} and \textit{targeted} advertising expose potentially sensitive and damaging information about users, also demonstrated in \cite{datta2014automated, rao2015they, book2015empirical}. There is a growing user awareness of privacy and a number of privacy initiatives, e.g., Apple's enabling of ad blockers in iOS9\footnote{\sloppy http://au.pcmag.com/mobile-operating-system/31341/opinion/apple-ios-9-ad-blocking-explained-and-why-its-a-ba} is representative of a move towards giving users greater control over the display of ads, although applicable only to browser based rather than to mobile targeted ads, however, this would greatly affect Google's services, since Google's services are now based on \textit{Web \& App} activity\footnote{My Google Activity: \url{https://myactivity.google.com/myactivity?otzr=1}}.

Hence, the purpose of \textit{targeted} advertising is to be able to protect user's privacy and effectively serve relevant ads to appropriate users, in particular, to enable private \textit{profiling} and \textit{targeted} ads without revealing user interests to the adverting companies or third party ad/tracking companies. Furthermore, an private \textit{billing} process to update the advertising network about the ads retrieved/clicked in a privacy preserving manner.


\subsection{Privacy attacks}
There are various kinds of privacy attacks, we mainly focus on three main categories of privacy attacks. Note that in all these scenarios, the user is not opposed to \textit{profiling} in general and is willing to receive services e.g., \textit{targeted} ads, on selected topics of interests, but does not wish for specific parts of their profile (\textit{attributes}), based on the usage of \textit{apps} (s)he considers private, to be known to the \textit{analytics} network or any other party, or to be used for personalised services.

\subsubsection{Unintended privacy loss} In this case, users voluntary provide personal information, e.g. to OSNs, or users authorize third-party services to access personal information, e.g. third-party library tracking in mobile \textit{apps}, however users may not be aware how the information is used and what are the potential privacy risks.

\subsubsection{Privacy leakage via cross-linking or de-anonymisation} The user profile is (legitimately) derived by the \textit{analytics} network ( e.g. \cite{tchen2014, ullah2020protecting, ullah2017enabling} focused on Google AdMob and Flurry) by cross-linking private information or via de-anonymisation. In the former case, the \textit{analytics} services aggregate user data from sources that supposedly come as a results of users (willingly) previously shared their data with various data owners for providing them personalised services. In the later case, the data owners release anonymised personal information or data sources that sell data to advertisers or data anonymised data freely available on various websites\footnote{Kaggle dataset: \url{https://www.kaggle.com/datasets}, Dataset Search: \url{https://datasetsearch.research.google.com/}.}. The anonymised data is used to leak privacy when attackers disclose the identity of the data owner by cross-linking to external data sources i.e. using background knowledge \cite{tchen2014}.

\subsubsection{Privacy leakage vis statistical inference} The statistical inference i.e., an \textit{indirect} attack over user privacy, that involves a third party profile users based on their behavior to provide personalised services e.g. the advertising systems e.g., Google or Flurry monitor the ad traffic \cite{tchen2014, ullah2014characterising} sent to mobile devices and infers the user profile based on their \textit{targeted} ads. The profiling attributes are sensitive to the users and are considered as private information e.g. political view, religious, sexual orientation, etc.

\subsection{Ad traffic analysis for evaluating privacy leakage}
Several works investigate the mobile targeted ads traffic primarily for the purpose of privacy and security concerns. The AdRisk \cite{grace2012unsafe}, an automated tool, analyse 100 \textit{ad libraries} and studies the potential security and privacy leakages of these libraries. The \textit{ad libraries} involve the resource permissions, permission probing and \texttt{JavaScript} linkages, and dynamic code loading. Parallel to this work, \cite{stevens2012investigating} examines various privacy vulnerabilities in the popular Android-based \textit{ad libraries}. They categorise the permissions required by ad libraries into \textit{optional}, \textit{required}, or \textit{un-acknowledged} and investigate privacy concerns such as how user's data is sent in ad requests. The authors in \cite{liu2019privacy} analyse the privacy policy for collecting \textit{in-app} data by \textit{apps} and study various information collected by the \textit{analytics libraries} integrated in mobile \textit{apps}.

Other works \cite{pearce2012addroid, shekhar2012adsplit} study the risks due to the lack of separate working mechanisms between Android \textit{apps} and ad libraries and propose methods for splitting their functionality. The authors in \cite{leontiadis2012don} monitor the flow of data between the ad services and 250K Android \textit{apps} and demonstrate that currently proposed privacy protecting mechanisms are not effective, since \textit{app} developers and ad companies do not show any concern about user's privacy. They propose a market-aware privacy-enabling framework with the intentions of achieving symmetry between developer's revenue and user's privacy. Another work \cite{book2013longitudinal} carried out a longitudinal study in the behaviour of Android \textit{ad libraries}, of 114K free \textit{apps}, concerning the permissions allocated to various \textit{ad libraries} over time. The authors found that over several years, the use of most of the permissions has increased over time raising privacy and security concerns.


There has been several other works, exploring the web advertisements in different ways i.e. form the monetary perspective \cite{aggarwal2009general, yan2009much}, from the perspective of privacy of information of users \cite{guha2009serving}, from privacy information leakage and to propose methods to protect user data \cite{krishnamurthy2009leakage, krishnamurthy2010privacy}, and the E-Commerce \cite{metwally2007detectives}. In similar way, a detailed analysis of the web ad networks from the perspective information communicated on network level, the network layer servers, and from the point of the content domains involved in such a system are investigated \cite{wang2011understanding}.

\subsection{Inference of private information}
In recent years, several works \cite{schwartz2013personality, kosinski2013private, goel2012does, hu2007demographic, schler2006effects, otterbacher2010inferring, mukherjee2010improving, bi2013inferring, ying2012demographic} have shown that it is possible to infer undisclosed private information of subscribers of online services such as age, gender, relationship status, etc. from their generated contents. The authors in \cite{schler2006effects} analysed the contents of 71K blogs at \url{blogger.com} and were able to accurately infer the gender and age of the bloggers. The authors were able to make their inferences by identifying certain unique features pertaining to an individual's writing style such as parts-of-speech, function words, hyper-links and content such as simple content words and the special classes of words taken from the handcrafted LIWC (Linguistic Inquiry and Word Count) \cite{pennebaker2001linguistic} categories.

Another study \cite{schwartz2013personality} has shown that the age demographics of Facebook users (both using \textit{apps} and browsers) can be predicted by analysing the language used in status update messages. Similar inferences have been made for IMDB users based on their movie reviews \cite{otterbacher2010inferring}. Another work \cite{bi2013inferring} predicts age, gender, religion, and political views of users from the queries using models trained from Facebook's `Like' feature. In \cite{goel2012does}, the authors analysed client-side browsing history of 250K users and were able to infer various personal attributes including age, gender, race, education and income. Furthermore, a number of studies \cite{zheleva2009join, he2006inferring, mislove2010you} have demonstrated that sensitive attributes of user populations in online social networks can be inferred based on their social links, group memberships and the privacy policy settings of their friends \cite{ryu2013curso}.

\subsection{User information extraction}
We experimentally evaluate \cite{tchen2014} how to extract user profiles from mobile \textit{analytics} services based on the device identifier of the target; this method was demonstrated using both Google \textit{analytics} and Flurry in the Android environment. Here the user profile, i.e. set of information collected or inferred by the \textit{analytics} services, consists of personally identifiable information such as, unique device ID, demographics, user interests inferred from the \textit{app} usage etc.

An crucial technique to extract user profiles from the \textit{analytics} services (we mainly target Google and Flurry \textit{analytics} services) is to first impersonate the victim's identity; then \textbf{\textit{Case 1 Google analytics}}: to fetch user profiles from a spoofed device, where the private user profile is simply shown by the Google service as an ads preference setting or \textbf{\textit{Case 2 Flurry analytics}}: to inject the target's identity into a controlled \textit{analytics} \textit{app}, which impacts those changes in the Flurry audience analysis report using which the adversary is able to extract user profile. Following, we first describe how to obtain and spoof a device's identity, subsequently, the user profile extraction for both cases of Google and Flurry is presented in detail.

\subsubsection{Information extraction via user profiles from Google} Android system allows users to view and manage their \textit{in-app} ads preferences\footnote{Access from \texttt{Google Settings} $\to$ \texttt{Ads} $\to$ \texttt{Ads by Google} $\to$  \texttt{Ads Settings}. It claims that Google's ad network shows ads on 2+million non-Google websites and \textit{apps}.}, e.g. to \textit{opt-out} or to \textit{update/delete} interests. This feature retrieves user profile from Google server which is identified by the advertising ID. As a consequence of the device identity spoofing, an adversary is able to access the victim's profile on a spoofed device.

We note that there are at least two possible ways to that an adversary can capture victims Android ID. First, an adversary can intercept the network communication, in order to capture the usage reporting messages sent by third-party tracking APIs, extract the device identifier, and to further use it for ongoing communication with the \textit{analytics} services. Note that it is very easy to monitor IDs of thousands of users in a public hotspots e.g. airport, hospital etc. Similarly, in a confined area, an adversary (e.g. an employer or a colleague) \textit{targeting} a particular individual can even associate the collected device ID to their target (e.g. employees or another colleague). During this privacy attack, we note that Google \textit{analytics library} prevents leakage of device identity by hashing the Android IDs; however it cannot stop other \textit{ad libraries} to transmit such information in plain text (which can be easily be mapped to Google's hashed device ID).

An alternative way, although may be more challenging in practice, is to obtain the target's device identifier from any application (controlled by the adversary) that logs and exports the device's identity information.


\subsubsection{Information extraction via user profiles from Flurry} We note that extracting user profiles from Flurry is more challenging since Flurry does not directly allow users to view or edit user's \textit{Interests profiles}. In fact, except the initial consent on the access of device resources, many smartphone users may not be aware of the Flurry's tracking activity.

\begin{figure}[h]
\begin{center}
\includegraphics[scale=0.8]{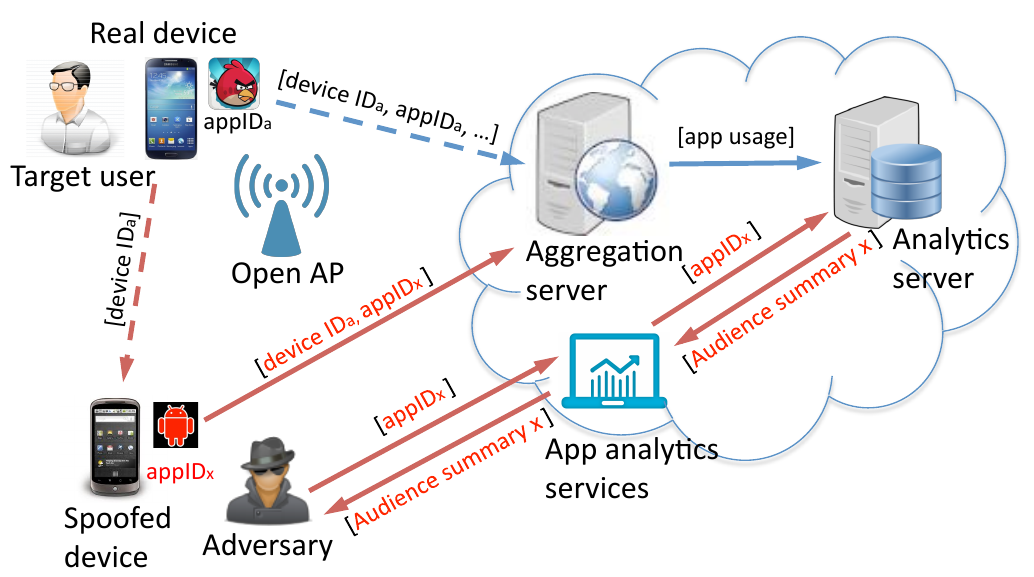}
\caption{Privacy leakage attack scenario \cite{tchen2014}.}
\label{fig:attack_privacy}
\end{center}
\end{figure}

Figure \ref{fig:attack_privacy} shows the basic operations of our profile extraction technique within the mobile advertising ecosystem. To compromise a user's private profile, an adversary spoofs the target device, identified by $deviceID_a$, using another Android device or an emulator. Following, the adversary uses a \textit{bespoke} \textit{app} with a (legitimate) $appID_x$, installed on the \textit{spoofed} device, to trigger a usage report message to Flurry. Accordingly, the \textit{analytics} service is manipulated into believing that $deviceID_a$ is using a new application tracked by the system. Consequently, all user related private information is made accessible to the adversary through audience analysis report of $appID_x$ in Flurry system.

An adversary can easily extract the corresponding statistics and link them to (legitimate) user once the audience report from Flurry targets a unique user. In addition, the adversary will be able to track and access all subsequent changes to the user profile at a later time. In our presented technique, since we do impersonate a particular target's device ID, we can easily associate the target to a `blank' Flurry-monitored application.

Alternatively, an adversary can derive an individual profile from an aggregated audience analysis report by monitoring report differences before and after a target ID has been spoofed (and as such has been added to the audience pool). Specifically, the adversary has to take a snapshot of the audience analysis report $P_t$ at time $t$, impersonates a target's identity within his controlled Flurry-tracked application, and then takes another snapshot of the audience analysis report at $P_{t+1}$. The target's profile is obtained by extracting the difference between $P_t$ and $P_{t+1}$, i.e. $\Delta(P_t,P_{t+1})$. However in practice, Flurry service updates profile attributes on a weekly basis which means it will take up to a week to extract a full profile per user.

Finally, the \textit{segment} feature provided by Flurry, the \textit{app} audience is further split by applying filters according to e.g. gender, age group and/or developer defined parameter values. This feature allows an adversary to isolate and extract user profiles in a more efficient way. For instance, a possible \textit{segment} filter can be `only show users who have Android ID value of $x$' which results in the audience profile containing only one particular user. The effectiveness of the attack are validated in two steps: \textbf{1}. We first validate that user's profile is the basis for  ads \textit{targeting}, by  showing  that  specific  profiles  will  consistently receive highly similar ads and conversely, that a difference  in  a  user's  profile  will  result  in  a  mobile  receiving dissimilar ads. \textbf{2}. We then perform the ad influence attack, i.e. we perturb selected profiles and demonstrate that the modified profiles indeed receive \textit{in-app} ads in  accordance with the profile modifications.

\subsection{Third-party privacy threats}
The third-party A\&A libraries have been examined in a number of works, such as \cite{grace2012unsafe, vallina2012breaking, han2012study, stevens2012investigating, enck2014taintdroid}, which contribute to the understanding of mobile tracking and collecting and disseminating personal information in current mobile networks. The information stored and generated by smartphones, such as call logs, emails, contact list, and GPS locations, is potentially highly sensitive and private to the users. Following, we discuss various means through which users' privacy is exposed.

\subsubsection{Third-party tracking}
Majority of privacy concerns of smartphone users are because of inadequate access control of resources within the smartphones e.g. Apple iOS and Android, employ fine-grained permission mechanisms to determine the resources that could be accessed by each application. However, smartphone applications rely on users to allow access to these permissions, where users are taking risks by permitting applications with malicious intentions to gain access to confidential data on smartphones \cite{ongtang2012semantically}. Similarly, privacy threats from collecting individual's online data (i.e. direct and inferred leakage), have been examined extensively in literature, e.g. \cite{mamais2019privacy, frik2020impact}, including third party ad tracking and visiting \cite{shuba2020nomoats, iqbal2020adgraph}.

Prior research works show the extent to which consumers are effectively tracked by a number of third parties and across multiple \textit{apps} \cite{razaghpanah2018apps}, mobile devices leaking \textit{PII} \cite{elsabagh2020firmscope, ren2016recon}, \textit{apps} accessing  user's  private and sensitive information through well defined APIs \cite{verderame2020reliability}, inference attacks based on monitoring ads \cite{tchen2014} and other data platform such as eXelate\footnote{\url{https://microsites.nielsen.com/daas-partners/partner/exelate/}}, BlueKai\footnote{\url{https://www.oracle.com/corporate/acquisitions/bluekai/}}, and AddThis\footnote{\url{https://www.addthis.com/}} that collect, enrich and resell cookies.

The authors in \cite{felt2012android} conducted a user survey and showed that minor number of users pay attention to granting access to permissions during installation and actually understand these permissions. Their results show that 42\% of participants were unaware of the existing permission mechanism, only 17\% of participant paid attention to permissions during \textit{apps} installation while only 3\% of participants fully understood meaning of permissions accessing particular resources. The authors in \cite{grace2012unsafe} evaluate potential privacy and security risks of information leakage in mobile advertisement by the embedded \textit{libraries} in mobile applications. They studied 100,000 Android \textit{apps} and identified 100 representative \textit{libraries} in 52.1\% of \textit{apps}. Their results show that the existing \textit{ad libraries} collect private information that may be used for legitimate \textit{targeting} purposes (i.e., the user location) while other data is harder to justify, such as the users call logs, phone number, browser bookmarks, or even the list of \textit{apps} installed on the phone. Additionally, they identify some \textit{libraries} that use unsafe mechanisms to directly fetch and run code from the Internet, which also leads to serious security risks. A number of works \cite{adrienne2011usenix, felt2011android, chan2012droidchecker}, identify the security risks on Android system by disassembling the applications and tracking the flow of various methods defined within various programmed classes.

There are several works to protect privacy by assisting users to manage permissions and resource access. The authors in \cite{enck2009lightweight} propose to check the \texttt{manifest}\footnote{Every Android \textit{app} contains the \textit{manifest} file that describes essential information about app, such as, \textit{app ID}, \textit{app name}, \textit{permission to use device resources used by an app e.g. contacts, camera, list of installed apps etc.}, \textit{hardware and software features the app requires} etc. \url{https://developer.android.com/guide/topics/manifest/manifest-intro}.} files of installed mobile \textit{apps} against the permission assignment policy and blocking those that request certain potentially unsafe permissions. The MockDroid \cite{beresford2011mockdroid} track the resource access and rewrites privacy-sensitive API calls to block information communicated outside the mobile phones. Similarly, the AppFence \cite{hornyack2011these} further improves this approach by adding taint-tracking, hence, allowing more refined permission policies.

\subsubsection{Re-identification of sensitive information}
Re-identification involves service personalisation based on pervasive spatial and temporal user information that have already been collected e.g. locations that users have already visited. The users are profiled and later on provided with additional offers based on their interests, such as, recommending on places to visit, or people to connect to. There have been a number of research works to identify users based on re-identification technique. For instance, the authors in \cite{golle2009anonymity} analyse U.S. Census data and show that on average, every 20 individuals from the dataset share same home or work locations while 5\% of people in dataset can be uniquely identified by home-work location pairs. Another related work \cite{zang2011anonymization} uniquely identifies US mobile phone users using generalisation technique by generalising the top $N$ homework location pairs. They use location information to derive quasi-identifiers for re-identification of users. Similarly, a number of research works e.g. \cite{mohammed2009walking, bonchi2011trajectory, shokri2011quantifying}, raise privacy issues in publishing sensitive information and focus on theoretical analysis of \textit{obfuscation} algorithms to protect user privacy.

\subsection{Quantifying privacy algorithms}
Quantifying privacy is an important and challenging task as it is important to evaluate the level of privacy protection achieved. It is difficult to formulate a generic metric for quantifying privacy that is applicable to different contexts and due to several types of privacy threats. It is also the different solutions i.e. specific techniques (not necessarily threats) that contain their unique privacy metrics, which are not cross-comparable.

For instance, the proposal for fulfilling the privacy requirements using $k$-anonymity, first proposed in \cite{samarati2001protecting}, requires that each equivalence class i.e. set of records that are indistinguishable from each other with respect to certain identifying attributes, must have a minimum of $k$ records \cite{sweeney2002k}. Another study \cite{machanavajjhala2007diversity} reveals that satisfying the privacy requirements for $k$-anonymity cannot always prevent attribute disclosures mainly for two reasons: First, an attacker can easily discover the sensitive attributes when there is minute diversity in the sensitive attributes, secondly, $k$-anonymity is not resistant to privacy attacks against the attackers that use background knowledge. They \cite{machanavajjhala2007diversity} proposes an $l$-diversity privacy protection mechanism against such attacks and evaluates its practicality both formally and using experiment evaluations. Another work \cite{li2007t} evaluates the limitation of $l$-diversity and proposes $t$-closeness, suggesting the distribution of sensitive attributes in an equivalence class must be close to the distribution of attributes in the overall data i.e. distance between two distributions should not be more than the $t$ threshold.

Besides, techniques based on crypto mechanisms, such as PIR, provide privacy protection, for the database present on \textit{single-server}, against the computational complexity \cite{aguilar2007lattice, chor1997computationally}, \textit{multiple-servers} for protecting privacy against colluding adversaries \cite{goldberg2007improving, henry2011practical, beimel2004reducing, gertner1998random, devet2012optimally}, or protection mechanisms \cite{devet2014best} against combined privacy attacks that are either computationally bounded evaluations or against colluding adversaries; these techniques are discussed in detail in Appendix \ref{pir-mechanisms}.

\section{Privacy in Mobile Ads: Solutions} \label{solutions}

\begin{figure*}[t]
\begin{center}
\includegraphics[scale=0.6]{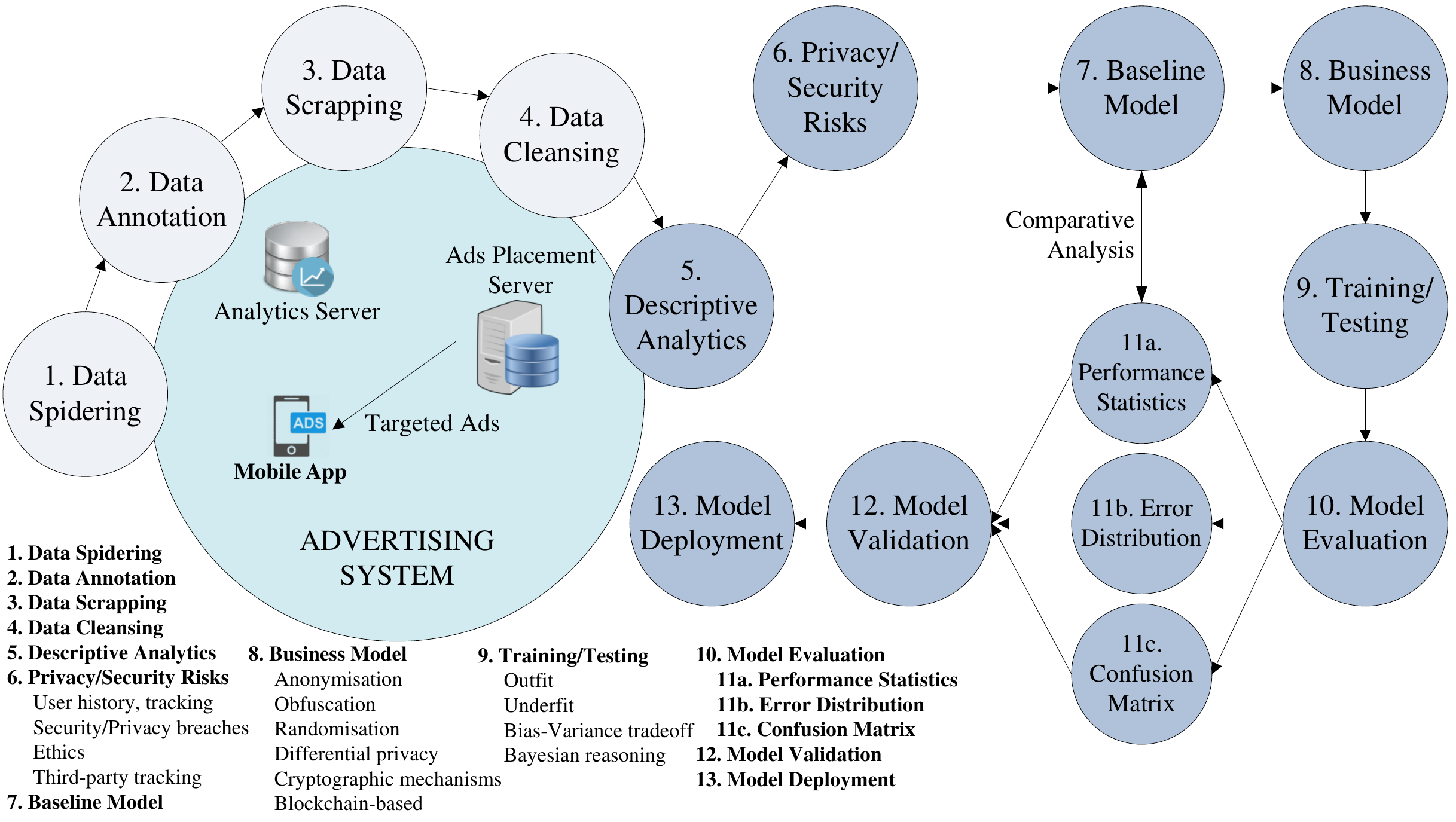}
\caption{Lifecycle of proposal for privacy-preserving advertising systems for both \textit{in-app} and \textit{in-browser} targeted ads.}
\label{privacy-preserving}
\end{center}
\end{figure*}

The \textit{direct} and \textit{indirect} (i.e., inferred) leakages of individuals' information have raised privacy concerns. A number of research works propose private \textit{profiling} (and advertising) systems \cite{toubiana2010adnostic, fredrikson2011repriv, guha2011privad, haddadi2010mobiad, chen2012towards, chen2013splitx}. These systems do not reveal either the users' activities or the user's interest profiles to the ad network. Various mechanisms are used to accomplish these goals: Adnostic \cite{toubiana2010adnostic}, Privad \cite{guha2011privad} and Re-priv \cite{fredrikson2011repriv} focus on \textit{targeting} users based on their browsing activities, and are implemented as browser extensions running the \textit{profiling} algorithms locally (in the user's browser). MobiAd \cite{haddadi2010mobiad} proposes a distributed approach, specifically aimed at mobile networks. The use of \textit{differential privacy} is advocated in \textit{Practical Distributed Differential Privacy} (PDDP) \cite{chen2012towards} and SplitX \cite{chen2013splitx}, where differentially private queries are conducted over distributed user data. All these works protect the full user profile and advocate the use of novel mechanisms that necessitate the re-design of some parts or all of the current advertising systems, although some (e.g., Adnostic) can operate in parallel with the existing systems. In addition, the works based on the use of noisy techniques like \textit{differential privacy}, to obfuscate user's preferences may result in a lower accuracy of \textit{targeted} ads (and correspondingly lower revenues), compared to the use of standard \textit{targeting} mechanisms.

Figure \ref{privacy-preserving} shows the lifecycle of proposal for privacy-preserving mobile/web advertising systems; specifically starting from data collection for evaluating privacy/security risks, baseline model and proposed business model for preserving user's privacy, finally model evaluation and its comparison with the baseline model. Various data scrapping elements, statistical measures and privacy preserving techniques are also shown in this figure.

An important thing in the development of private advertising system is that the consumers' trust in privacy of mobile advertising is positively related to their willingness to accept mobile advertising \cite{tsang2004consumer, merisavo2007empirical}. The AdChoices\footnote{\url{https://optout.aboutads.info/?c=2&lang=EN}} program (a self-regulation program implemented by the American ad industry), states that consumer could \textit{opt-out} of \textit{targeted} advertising via online choices to control ads from other networks. However, another study \cite{johnson2020consumer} examines that the \textit{opt-out} users cause 52\% less revenue (and hence presents less relevant ads and lower click through rates) than those users who allow \textit{targeted} advertising. In addition, the authors noted that these ad impressions were only requested by 0.23\% of American consumers.

\subsection{Private ad ecosystems}
There are a number of generic privacy preserving solutions proposed to address the negative impact of ads \textit{targeting}. Anonymity solutions for web browsing include the use of Tor \cite{dingledine2004tor}, or disabling the use of cookies \cite{aggarwal2010analysis}. These accomplish the goal of preventing user tracking, however, they also prevent any user (profile based)  service personalisation, that may actually be a desirable feature for many users despite their privacy concerns.

Research proposals to enable privacy preserving advertising have been more focused on web browsing, as the dominant advertising media e.g., \cite{rafieian2020targeting, Akkus:2012, guha2011privad, toubiana2010adnostic, chen2013splitx}, propose to use locally derived user profiles. In particular, Privad \cite{guha2011privad} and Adnostic \cite{toubiana2010adnostic} use the approach of downloading a wide range of ads from the ad network and locally (in the browser or on the mobile device) selecting ads that match the user's profile. On the other hand, there are a smaller number of works address privacy for mobile advertising, with representative works e.g., \cite{ullah2020privacy, ullah2020protecting, ullah2017enabling, ullah2014profileguard, haddadi2010mobiad, backes2012obliviad, Hardt:2012}, suggest the \textit{app}-based user \textit{profiling}, stored locally on mobile device. The \cite{ullah2017enabling} is based on various mechanisms of PIR and it complements the existing advertising system and is conceptually closest to \cite{backes2012obliviad}, which uses Oblivious RAM (ORAM) to perform Private Information Retrieval (PIR) on a secure coprocessor hardware. However, unlike our solution it relies on specific (secure) hardware to enable PIR, which may limit its applicability in a general setting.


\subsection{Data masking, anonymisation, obfuscation and randomisation}

There are several privacy protection techniques, such as techniques based on \textit{anonymisation} e.g. encrypting or removing \textit{PII}, \textit{proxy-based} solutions, \textit{k-anonymity} i.e. \textit{generalisation} and \textit{suppression}, \textit{obfuscation} (making the message confusing, willfully ambiguous, or harder to understand), mechanisms based on \textit{differential privacy} i.e. maximising the accuracy of queries from statistical databases while minimising the chances of identifying its records, \textit{crypto-based} techniques such as \textit{private information retrieval} (PIR) and \textit{blockchain-based} solutions. Following we present various privacy-preserving advertising systems based on these different techniques.

\subsubsection{Anonymisation}
The simplest and most straightforward way to \textit{anonymise} data includes masking or removing data fields (attributes) that comprise \textit{PII}. These include direct identifiers like names and addresses, and quasi-identifiers (QIDs) such as gender and zip code, or an IP address; the later can be used to uniquely identify individuals. It is assumed that the remainder of the information is not identifying and therefore not a threat to privacy (although it contains information about individuals, e.g. their interests, shopping patterns, etc.). A second approach is to generalise QIDs, e.g., by grouping them into a higher hierarchical category (e.g., locations into post codes); this can also be accomplished according to specified \textit{generalisation} rules. \textit{Anonymisation} mechanisms that deal with selected QIDs according to pre-determined rules include $k$-anonymity \cite{samarati1998generalizing} and it's variants like $l$-diversity \cite{machanavajjhala2007diversity} and $t$-closeness \cite{li2007t}. These, in their simplest form, $k$-anonymity (detailed discussion over $k$-anonymity is given in Appendix \ref{k-anonymity}), modifies (\textit{generalise}) individual user records so that they can be grouped into identical (and therefore indistinguishable) groups of $k$, or additionally apply more complex rules ($l$-diversity and $t$-closeness).

A number of proposals advocate the use of locally (either in the browser of the mobile device) derived user profiles, where user's interests are \textit{generalised} and/or partially removed (according to user's privacy preferences), before being forwarded to the server or an intermediary that selected the appropriate ads to be forwarded to the clients. In the context of \textit{targeted} advertising, the removal of direct identifiers includes user IDs (replacing them with temporary IDs) or mechanisms to hide used network address (e.g., using TOR \cite{dingledine2004tor}). However, if only the most obvious \textit{anonymisation} is applied without introducing additional (\textit{profiling} and \textit{targeting} oriented) features, the ad networks ecosystem would be effectively disabled. Therefore, we only mention representative solutions from this category and concentrate on the privacy-preserving mechanisms that enable \textit{targeted} ads.

The privacy requirements are also, in a number of prior works, considered in parallel with achieving bandwidth efficiency for ad delivery, by using caching mechanisms \cite{guha2011privad, khan2013cameo, haddadi2010mobiad}. Furthermore, such techniques have been demonstrated to be vulnerable to composition attacks \cite{ganta2008composition}, and can be reversed (with individual users identified) when auxiliary information is available (e.g. from online social networks or other publicly available sources) \cite{sweeney2000simple, coull2007playing}.

In Adnostic \cite{toubiana2010adnostic}, each time a webpage (containing ads) is visited by the user; the client software receives a set of generic ads, randomly chosen by the broker. The most appropriate ads are then selected locally, by the client, for presentation to the user; this is based on the locally stored user profile. We have categorised this work as a \textit{generalisation} mechanism as the served ads are generic (non-personalised), although it could arguably be considered under the \textit{randomisation} techniques. We note that in \cite{toubiana2010adnostic} the user's privacy (visited pages or ad clicks) is not protected from the broker.

In Privad \cite{guha2009privad, guha2011privad}, a local, (detailed) user profile is generated by the Privad client and then \textit{generalised} before sending to the ads broker in the process of requesting (broadly) relevant ads. All communication with the broker is done through the dealer, which effectively performs the functions of an \textit{anonymising} proxy; the additional protection is delivered by encrypting all traffic, this protecting user's privacy from the dealer. The proposed system also includes monitoring of the client software to detect whether any information is sent to the broker using, e.g., a covert channel. Similarly, in MobiAd \cite{haddadi2010mobiad}, the authors propose a combination of peer-to-peer mechanisms that aggregates information from users and only presents the aggregate (\textit{generalised} activity) to the ad provider, for both ad impressions and clicks. Caching is utilised to improve efficiency and Delay tolerant networking for forwarding the information to the ad network. Similarly, another work \cite{artail2015privacy} proposes combining of users interests via an ad-hoc network, before sending them to the ad server.

Additionally, some system proposals \cite{hardt2012privacy} advocate the use of \textit{anonymisation} techniques ($l$-diversity) in the \textit{targeting} stage, where the ads are distributed to users, while utilising alternative mechanisms for \textit{profiling}, learning and statistics gathering.


\subsubsection{Obfuscation}
\textit{Obfuscation} is the process of obscuring the intended meaning of the data or communication by making the message difficult to understand.

In the scenario of an advertising system, recall that the user privacy is mainly breached for their \textit{context} i.e. specific use of mobile \textit{apps} from an \textit{app} category, and their profiling \textit{interests} along with the ads targeting based on these interests. Hence, an important focus in implementing such mechanisms is to \textit{obfuscate} specific profiling attributes that are selected as private (i.e. the attributes that the analytics companies may use for interest-based advertisements) and the categories of installed \textit{apps}. For instance, the user may not wish the categories of gaming or porn to be included in their profile, as these would reflect heavy use of corresponding (gaming and porn) \textit{apps}. The \textit{obfuscation} scenarios can be based on similar (obfuscating) \textit{apps} or similar profiling attributes or interests customised to user's profile \cite{ullah2020protecting} or randomly chosen \textit{apps/interests} from non-private categories. An important factor is to take into consideration the extra (communication, battery, processing, usage airtime) overhead while implementing \textit{obfuscation} mechanisms, following, it needs present jointly optimised framework that is cost effective and preserves user privacy for profiling, temporal \textit{apps} usage behavioral patterns and interest-based ads targeting.

A recent work \cite{wermke2018large} carries out a large scale investigation of \textit{obfuscation} use where authors analyse 1.7 million free Android \textit{apps} from Google Play Store to detect various \textit{obfuscation} techniques, finding that only 24.92\% of \textit{apps} are obfuscated by the developer. There are several \textit{obfuscation} mechanisms for protecting private information, such as the \textit{obfuscation} method presented in \cite{weinsberg2012blurme} that evaluates different classifiers and \textit{obfuscation} methods including greedy, sampled and random choices of obfuscating items. They evaluate the impact of \textit{obfuscation}, assuming prior knowledge of the classifiers used for the inference attacks, on the utility of recommendations in a movie recommender system. A practical approach to achieving privacy \cite{salamatian2013hide}, which is based on the theoretical framework presented in \cite{du2012privacy}, is to distort the view of the data before making it publicly available while guaranteeing the utility of the data. Similarly, \cite{li2007protecting} proposes an algorithm for publishing partial data that is safe against the malicious attacks where an adversary can do the inference attacks using association rule in publicly published data.

Another work, `ProfileGuard' \cite{ullah2014profileguard} and its extension \cite{ullah2020protecting} propose an \textit{app}-based profile \textit{obfuscation} mechanism with the objective of eliminating the dominance of private interest categories (i.e. the prevailing private interest categories present in a user profile). The authors provide insights to Google AdMob \textit{profiling} rules, such as showing how individual \textit{apps} map to user's interests within their profile in a deterministic way and that AdMob requires a certain level of activity to build a \textit{stable} user profile. These works use a wide-range of experimental evaluation of Android \textit{apps} and suggest various \textit{obfuscation} mechanisms e.g. \textit{similarity} with user's existing \textit{apps}, \textit{bespoke} (customised to profile \textit{obfuscation}) and \textit{bespoke++} (\textit{resource-aware}) strategies. Furthermore, the authors also implement a POC `ProfileGuard' \textit{app} to demonstrate the feasibility of an automated \textit{obfuscation} mechanism.

Following, we provide an overview of prior work in both \textit{randomisation} (generic noisy techniques) and \textit{differentially private} mechanisms.

\subsubsection{Randomisation}
In the \textit{randomisation} methods, noise is added to distort user's data. Noise can either be added to data values (e.g., movie ratings or location GPS coordinates), or, more applicable to \textit{profiling} and user \textit{targeting}, noise is in the form of new data (e.g., additional websites that the user would not have visited normally are generated by a browser extension \cite{howe2009trackmenot}), added in order to mask the true vales of the records (browsing history). We note that \cite{howe2009trackmenot} protects the privacy of user's browsing interests but does not allow (privacy preserving) \textit{profiling} or selection of appropriate \textit{targeted} ads.

The idea behind noise addition is that specific information about user's activities can no longer be recovered, while the aggregate data still contains sufficient statistical accuracy so that it can be useful for analysis (e.g., of trends).
A large body of research work focuses on generic noisy techniques e.g. \cite{agrawal2000privacy} proposed the approach of adding random values to data, generated independently of the data itself, from a known e.g., the uniform distribution. Subsequent publications (e.g., \cite{evfimievski2003limiting}) improve the initial technique, however other research work \cite{kargupta2003privacy} has identified the shortcomings of this approach, where the added noise may be removed by data analysis and the original data (values) recovered.


A novel noisy technique for privacy preserving personalisation of web searches was also recently proposed \cite{mor2015bloom}. In this work, the authors use `Bloom' cookies that comprise a noisy version of the locally derived profile. This version is generated by using Bloom filters \cite{bloom1970space}, an efficient data structure; they evaluate the privacy versus personalisation trade-off.

\subsection{Differential privacy}

The concept of \textit{differential privacy}\footnote{A C++ implementation of \textit{differential privacy} library can be found at \url{https://github.com/google/differential-privacy}.} was introduced in \cite{dwork2006calibrating}, a mathematical definition for the privacy loss associated with any released data or \textit{transcript} drawn from a database. Two datasets $D_1$ and $D_2$ differ in at most one element given that one dataset is the subset of the other with larger database contains only one additional row e.g. $D_2$ can be obtained from $D_1$ by adding or removing a single user. Hence, a \textit{randomised} function $K$ gives \textit{differential privacy} for the two data sets $D_1$ and $D_2$ as: \({{\mathop{\rm P}\nolimits} _r}\left[ {K\left( {{D_1}} \right) \in S} \right] \le \exp \left( \varepsilon  \right) \times {{\mathop{\rm P}\nolimits} _r}\left[ {K\left( {{D_2}} \right) \in S} \right]\). We refer readers to \cite{dwork2014algorithmic} for deeper understanding of \textit{differential privacy} and its algorithms.

\textit{Differential privacy} is vastly used in the literature for \textit{anonymisation} e.g. a recent initiative to address the privacy concerns by recommending usage of \textit{differential privacy} \cite{cho2020contact} to illustrate some of the short-comings of direct contact-tracing systems. Google has recently published a \textit{Google COVID-19 Community Mobility Reports}\footnote{A publicly available resource to see how your community is moving around differently due to COVID-19: \url{http://google.com/covid19/mobility}} to help public health authorities understand the mobility trends over time across different categories of places, such as retail, recreation, groceries etc., in response to imposed policies aimed at combating COVID-19 pandemic. The authors in \cite{yan2020differential} use \textit{differential privacy} to publish statistical information of two-dimensional location data to ensure location privacy. Other works, such as \cite{zhang2014towards, zhang2016privtree}, partition data dimensions to minimise the amount of noise, and in order to achieve higher privacy accuracy, by using \textit{differential privacy} in response to the given set of queries.

\textit{Differential privacy} \cite{dwork2006differential} work has, in recent years, resulted in a number of system works that advocate the practicality of this, previously predominantly theoretical research field. The authors in \cite{chen2012towards} propose a system for \textit{differentially private} statistical queries by a data aggregator, over distributed users data. A proxy (assumed to be \textit{honest-but-curious}) is placed between the analyst (aggregator) and the clients and secure communications including authentication and traffic confidentiality  are accomplished using TLS \cite{dierks2008transport}. The authors also use a cryptography solution to provide additional privacy guarantees.The SplitX system \cite{chen2013splitx} also provides \textit{differential privacy} guarantees and relies on intermediate nodes, which forward and process the messages between the client that locally stores their (own) data and the data aggregator. Further examples include works proposing the use of distributed \textit{differential privacy} \cite{rastogi2010differentially} and \cite{shi2011privacy}.

\subsection{Cryptographic mechanisms} A number of different cryptographic mechanisms have been proposed in the context of \textit{profiling} and \textit{targeted} advertising or, more broadly, search engines and recommender systems. These include: Private Information Retrieval (PIR), Homomorphic encryption, Multi-party Computing (MPC), Blockchain based solutions.

\subsubsection{Private Information Retrieval (PIR)}
Private Information retrieval (PIR) \cite{chor1997computationally, kushilevitz1997replication, goldberg2007improving, Chor:1995:PIR:795662.796270, chor1997private, devet2014best}, is the ability to query a database successfully without the database server discovering which record(s) of the database was retrieved or the user was interested in. Detailed discussion over various PIR mechanisms along with their comparison is given in Appendix \ref{pir-mechanisms}.

The ObliviAd proposal \cite{backes2012obliviad} uses a PIR solution based on bespoke hardware (secure coprocessor), which enables on-the-fly retrieval of ads. The authors propose the use of Oblivious RAM (ORAM) model, where the processor is a ``black box'', with all internal operations, storage and processor state being unobservable externally.  ORAM storage data structure comprises of entries that include a combination of keyword and a corresponding ad (multiple ads result in multiple entries). The accounting and \textit{billing} are secured via the use of using electronic tokens (and mixing \cite{chaum1981untraceable, desmedt2000break}). More generally, a system that enables private e-commerce using PIR was investigated in \cite{henry2011practical}, with tiered pricing with record level granularity supported via the use of the proposed Priced Symmetric PIR (PS-PIR) scheme. Multiple sellers and distributed accounting and \textit{billing} are also supported by the system.

Additionally, cryptographic solutions can be used to provide part of the system functionality. They are commonly used in conjunction with \textit{obfuscation}, e.g., in \cite{rastogi2010differentially, shi2011privacy} or \textit{generalisation} \cite{toubiana2010adnostic}.

\subsubsection{Zero Knowledge Proof (ZKP) and Mixing}
zero knowledge proofs \cite{boudot2000efficient, schnorr1990efficient, brands2000rethinking, camenisch1999proving} and \textit{mixing} \cite{ghaderi2010towards} are commonly used as components of the privacy solutions. ZKP is a cryptographic commitment scheme by which one party (the \textit{prover}) can prove to another party (the \textit{verifier}) that they know a value $x$, without conveying any information apart from the fact that they know the value $x$. An example of \textit{Mixing}, called \textit{mixnet} \cite{chaum1981untraceable}, based  on  cryptography  and  permutation, was introduced to achieve anonymity in network communication. It creates a hard-to-trace communication by using a chain of proxy servers, called \textit{mixes}, which takes messages from multiple senders, shuffle, and send them back in random order to the destination, hence, breaking the link between source and destination and making it harder for eavesdroppers to trace end-to-end communications. A number of robust, threshold mix networks have appeared in the literature \cite{abe1998universally, piotrowska2020low, abe1999mix, desmedt2000break, jakobsson1998practical, jakobsson1999millimix, mitomo2000attack}.

Chen et al. \cite{chen2012towards} uses cryptographic mechanism to combine client-provided data (modified in accordance with \textit{differential privacy}). They utilise a probabilistic Goldwasser-Micali cryptosystem \cite{goldreich1987play}. In their subsequent work \cite{chen2013splitx}, the authors  use an XOR-based crypto-mechanism to provide both anonymity and unlinkability to analysis (queries) of \textit{differentially private} data distributed on user's devices (clients). A cryptography technique, \textit{mixing} \cite{chaum1981untraceable, desmedt2000break} is also commonly used as part of \textit{anonymisation} \cite{juels2001targeted, backes2012obliviad}, where \textit{mix} servers are used as intermediaries that permute (and re-encrypt) the input.

\subsubsection{Homomorphic encryption}
Homomorphic encryption \cite{yi2014homomorphic} is a form of encryption that allows specific types of computations to be carried out on ciphertext, without decrypting it first, and generates an encrypted result that, when decrypted, matches the result of operations performed on the plaintext.

Adnostic \cite{toubiana2010adnostic} uses a combination of homomorphic encryption and zero-knowledge proof mechanisms to enable accounting and \textit{billing} in the advertising system in a (for the user) privacy preserving way. Effectively, the user is protected as neither the publisher (website that includes the ads) or the advertisers (that own the ads) have knowledge about which users viewed specific ads. The authors in \cite{rastogi2010differentially} also combine \textit{differential privacy} with a homomorphic cryptosystem, to achieve privacy in a more generic setting of private data aggregation of distributed data. Similarly, Shi et al. \cite{shi2011privacy} also use a version of homomorphic techniques to enable private computing of sums based on distributed time-series data by a non-trusted aggregator.

The authors in \cite{erkin2012generating} presents privacy-preserving recommendations using partially homomorphic encryption (PHE) along with secure multi-party computation protocols. Specifically, user's private data encrypted via PHE, this way the recommender cannot use their original data while still being able to generate private recommendation, is uploaded to the recommender system; following the recommender runs a cryptographic protocol offline with a third party to generate personalised recommendations. This proposal also achieves good performance by lowering the processing and communication overheads by borrowing high cryptographic computations from third-party systems. Similarly, \cite{badsha2016practical} proposes a recommendation system based on the ElGamal cryptosystem (i.e. a kind of PHE), where all users actively collaborate with recommender server privately generate recommendations for a target user. Another work \cite{badsha2017privacy} relies on Boneh-Goh-Nissim (BGN) homomorphic cryptosystem that adopts an additional isolated recommender server that assists users in decrypting ciphertexts whenever necessary, hence, actively interact with both recommendation and additional servers.

\subsubsection{Multi-Party Computing (MPC)}
MPC \cite{cramer2005multiparty} is a set of cryptographic methods that allow private computing (of selected mathematical functions) on data from multiple, distributed, parties, without exposing any of the input data. The formal guarantees provided by MPC relate to both data confidentiality and the correctness of the computed result.

A web-based advertising system was first proposed by Juels \cite{juels2001targeted} , where they use multi-party \textit{information-theoretic} (threshold) PIR  in an \textit{honest-but-curious multi-server} architecture. Central to their system is the choice of a negotiant function, that is used by the advertiser to select ads, starting from a user's profile - the authors describe both a semi-private and a fully private \textit{information-theoretic} (threshold) PIR  in an \textit{honest-but-curious multi-server} architecture. They evaluate the benefits of both alternatives in regards to security, computational cost and communication overheads. In addition, in one of our previous works \cite{ullah2017enabling}, our motivation for using \textit{information-theoretic} (threshold) PIR for mobile private advertising system, rather than other solutions, e.g., Oblivious Transfer \cite{chu2008efficient, naor1999oblivious}, is the lower communication and computation overheads of such schemes. 

\begin{figure*}[h]
\begin{center}
\includegraphics[scale=0.5]{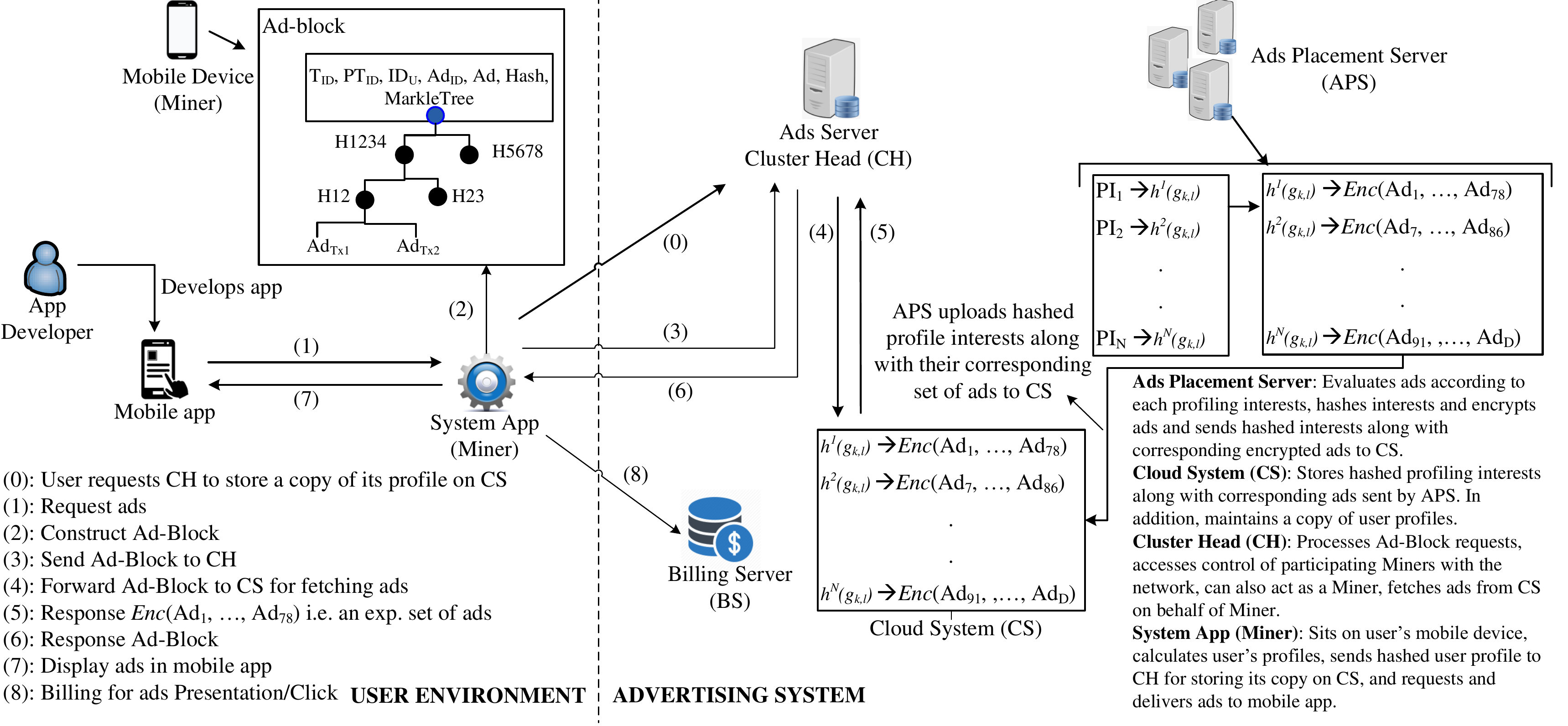}
\caption{A framework for secure user \textit{profiling} and Blockchain-based \textit{targeted} advertising system for \textit{in-app} mobile ads \cite{ullah2020privacy}. Description of various operation redirections (left side) and advertising entities (right side) is also give in this figure.}
\label{ad-system-blockchain}
\end{center}
\end{figure*}

\subsection{Blockchain-based advertising systems}
Blockchain is a fault-tolerant distributed system based on a distributed ledger of transactions, shared across the participating entities, and provides auditable transitions \cite{kosba2016hawk}, where the transactions are verified by participating entities within operating network. A blockchain is unalterable i.e. once recorded, the data in any block cannot be changed without altering of all the subsequent blocks; hence, it may be considered secure by design with high Byzantine fault tolerance e.g., one quarter of the participating nodes can be faulty but the overall system continues to operate normally.

Among the participating entities in a blockchain-based network; the \textit{Miner} is a special node responsible for generating transactions, adding them to the pool of pending transactions and organizing into a \textit{block} once the size of transactions reaches a specific \textit{block size}. The process of adding a new block to the Blockchain is referred to as \textit{mining} and follows a consensus algorithm, such as Proof of Work (POW) \cite{vukolic2015quest} and Proof of Stake (POS) \cite{wood2014ethereum}, which ensures the security of Blockchain against malicious (\textit{Miner}) users. The participating entities use the \textit{Public-Private Key} pair that is used to achieve the anonymity \cite{dorri2017blockchain}. Among various salient features of Blockchain, i.e. irreversible, auditable, updated near real-time, chronological and timestamp, which, in addition, disregards the need of a central controlling authority; thus making it a perfect choice for restricting communication between the mobile \textit{apps} and the analytics/ad companies and keeping individual's privacy.

Blockchain \cite{nakamoto2019bitcoin} has numerous applications and has been widely used, e.g. IoT \cite{dedeoglu2020blockchain}, Bid Data \cite{yang2020blockchain}, Healthcare \cite{tandon2020blockchain}, Banking and finance \cite{chen2020blockchain} etc. Blockchain has become a new foundation for decentralised business models, hence in the environment of advertising platform, made it a perfect choice for restricting communication between mobile \textit{apps} (which is potentially a big source of private data leakage) and the ad/analytics companies and keeping individual's privacy.

To our knowledge, we note that there are very limited works available for Blockchain-based mobile \textit{targeted} ads in the literature e.g. the \cite{gu2018secure} presents a decentralised \textit{targeted} mobile coupon delivery scheme based on Blockchain. The authors in this work match the behavioral profiles that satisfy the criteria for \textit{targeting} profile, defined by the vendor, with relevant advertisements. However, we note that this framework does not include all the components of an advertising system including user profiles construction, detailed structure of various Blockchain-based transactions and operations, or other entities such as \textit{Miner} and the \textit{billing} process. Our recent work, \textit{AdBlock} \cite{ullah2020privacy}, presents a detailed framework (in addition to Android-based POC implementation i.e. a \textit{Bespoke Miner}) for privacy preserving user \textit{profiling}, privately requesting ads, the \textit{billing} mechanisms for presented and clicked ads, mechanism for uploading ads to the cloud, various types of transactions to enable advertising operations in Blockchain-based network, and methods for \textit{access policy} for accessing various resources, such as accessing ads, storing mobile user profiles etc. This framework is parented in Figure \ref{ad-system-blockchain}. We further experimentally evaluate its applicability by implementing various critical components: evaluating user profiles, implementing \textit{access policies}, encryption and decryption of user profiles. We observe that the processing delays with various operations evaluate to an acceptable amount of processing time as that of the currently implemented ad systems, also verified in \cite{ullah2017enabling}.

\begin{table*}[t]
\begin{center} \scalebox{0.9}{
{\begin{tabular}{|l|l|l|l|c|}
\hline
\textbf{Ref} & \textbf{Architecture} & \textbf{Mechanism} & \textbf{Deployment} & \textbf{Domain}\tabularnewline
\hline
Privad \cite{guha2011privad} & 3rd-party anonymising proxy & Crypto  & Browser add-on & \multirow{6}{*}{Web}\tabularnewline
\cline{1-4}
Adnostic \cite{toubiana2010adnostic} & Complements to existing sys & Crypto billing & Firefox extension & \tabularnewline
\cline{1-4}
PASTE \cite{rastogi2010differentially} & Untrusted third party & Fourier Perturbation Algo & Browser add-on & \tabularnewline
\cline{1-4}
\cite{freudiger2009towards} & Cookie management & User preference & \multirow{2}{*}{Standalone} & \tabularnewline
\cline{1-3}
\cite{akkus2012non} & Anonymising proxy & Differential privacy &  & \tabularnewline
\cline{1-4}
DNT \cite{dnt2009christopher}\footnote{} & \multirow{2}{*}{Delay Tolerant Network} & HTTP header & Browser side & \tabularnewline
\cline{1-1} \cline{3-5}
MobiAd \cite{haddadi2010mobiad} &  & Encryption & Mobile phone & \multirow{8}{*}{Mobile}\tabularnewline
\cline{1-4}
ObliviAd \cite{backes2012obliviad} & \multirow{6}{*}{Complements existing sys} & Crypto-based & \multirow{6}{*}{Client/Server sides} & \tabularnewline
\cline{1-1} \cline{3-3}
\cite{Hardt:2012} &  & Differential privacy &  & \tabularnewline
\cline{1-1} \cline{3-3}
SplitX \cite{chen2013splitx} &  & XOR-based encryption &  & \tabularnewline
\cline{1-1} \cline{3-3}
CAMEO \cite{khan2013cameo} &  & Context prediction &  & \tabularnewline
\cline{1-1} \cline{3-3}
ProfileGuard \cite{ullah2014profileguard, ullah2020protecting} &  & Profile Obfuscation &  & \tabularnewline
\cline{1-1} \cline{3-3}
\cite{gu2018secure} &  & \multirow{2}{*}{Blockchain} &  & \tabularnewline
\cline{1-1}
AdBlock \cite{ullah2020privacy} &  & &  & \tabularnewline
\cline{1-4}
\cite{ullah2017enabling} & Autonomous system & Crypto-based & Standalone & \tabularnewline
\hline
\end{tabular}}}
 \end{center}\caption{Summary of the \textit{in-browser} and \textit{in-app} advertising systems. \label{table:sysarchs}}
\end{table*}



Summary of various privacy preserving approaches, in terms of \textit{architecture}, \textit{mechanism}, \textit{deployment} and \textit{app domain}, for both \textit{in-browser} and mobile advertising systems is given in Table \ref{table:sysarchs}.

\begin{table*}[h]
\begin{center} \scalebox{0.65}{
{
\begin{tabular}{|l|c|c|c|c|c|c|c|}
\hline
\multirow{2}{*}{\textbf{Parameters}} & \textbf{Differential} & \multicolumn{2}{c|}{\textbf{Obfuscation}} & \textbf{Cryptographic} & \multirow{2}{*}{\textbf{Randomisation}} & \textbf{Blockchain} & \multirow{2}{*}{\textbf{Anonymisation}}\tabularnewline
\cline{3-4}
 & \textbf{Privacy} & \textbf{App-based} & \textbf{Profile-based} & \textbf{mechanisms} &  & \textbf{solutions} & \tabularnewline
\hline
\hline
Apps usage behavioral & \multirow{2}{*}{No guarantee} & \multirow{2}{*}{Guaranteed} & \multirow{2}{*}{No guarantee} & \multirow{2}{*}{No guarantee} & \multirow{2}{*}{No guarantee} & \multirow{2}{*}{No guarantee} & \multirow{2}{*}{No guarantee}\tabularnewline
privacy &  &  &  &  &  &  & \tabularnewline
\hline
\multirow{2}{*}{Profiling privacy} & Yes & Yes & Yes & \multirow{2}{*}{Yes} & Yes & \multirow{2}{*}{Yes} & \multirow{2}{*}{Yes}\tabularnewline
 & Low & (Low to high) & (Low to high) &  & (Low to high) &  & \tabularnewline
\hline
Indirect privacy exposure & \multirow{2}{*}{Yes} & \multirow{2}{*}{Yes} & \multirow{2}{*}{Yes} & \multirow{2}{*}{No} & \multirow{2}{*}{Yes} & \multirow{2}{*}{No} & \multirow{2}{*}{Yes}\tabularnewline
from targeted ads &  &  &  &  &  &  & \tabularnewline
\hline
Cost of achieving user & \multirow{2}{*}{Low} & \multirow{2}{*}{High} & \multirow{2}{*}{Low} & \multirow{2}{*}{High} & \multirow{2}{*}{Low} & \multirow{2}{*}{High} & \multirow{2}{*}{Low}\tabularnewline
privacy &  &  &  &  &  &  & \tabularnewline
\hline
\multirow{2}{*}{Targeted ads} & \multirow{2}{*}{Yes (Lower)} & Lower to no relevant & Lower to no relevant & \multirow{2}{*}{Yes} & Lower to no relevant & \multirow{2}{*}{Yes} & \multirow{2}{*}{Yes}\tabularnewline
 &  & ads (adjustable) & ads (adjustable) &  & ads (adjustable) &  & \tabularnewline
\hline
Tradeoff b/w privacy and & \multirow{2}{*}{No} & \multirow{2}{*}{Yes} & \multirow{2}{*}{Yes} & \multirow{2}{*}{No} & \multirow{2}{*}{Yes} & \multirow{2}{*}{No} & \multirow{2}{*}{No}\tabularnewline
targeted ads &  &  &  &  &  &  & \tabularnewline
\hline
Impact over billing for & \multirow{2}{*}{Yes} & \multirow{2}{*}{Yes} & \multirow{2}{*}{Yes} & \multirow{2}{*}{No} & \multirow{2}{*}{Yes} & \multirow{2}{*}{No} & \multirow{2}{*}{No}\tabularnewline
targeted ads &  &  &  &  &  &  & \tabularnewline
\hline
\end{tabular}
}}
 \end{center} \caption{Comparison of various privacy protection mechanisms for various important parameters applicable in an advertising system. \label{privacy-comparision}}
\end{table*}

provides a hypothetical comparison of various privacy protection mechanisms using different parameters, evaluated in our proposed framework.

\subsection{Comparison of various privacy protection mechanisms proposed in an ad system}
Table \ref{privacy-comparision} provides a hypothetical comparison of various privacy protection mechanisms for various important parameters applicable in an advertising system, e.g., \textit{Apps} or \textit{Interest} profiling privacy, cost of achieving user privacy etc. We plan to carry out a comprehensive study over these parameters for above (presented in Table \ref{privacy-comparision}) privacy protection mechanisms in the future, in order to validate/invalidate our hypotheses.

It can be observed that the \textit{Obfuscation}-based mechanisms can guarantee user's `apps usage behavior privacy' (as evident in \cite{ullah2014profileguard, ullah2020protecting}) at the expense of installing and running a number of mobile \textit{apps}, similarly, the `cost' of achieving user privacy with \textit{Blockchain}-based solution is quite high due to its operational complexity \cite{gu2018secure, ullah2020privacy}. An important parameter is `impact over \textit{targeted} ads' as a results of achieving user privacy with various techniques e.g. \textit{Crypto-based} techniques (such as PIR),  \textit{Blockchain} and \textit{Anonymisation} techniques will have no impact over \textit{targeted} ads, alternatively, the \textit{Differential privacy}, \textit{Obfuscation} and \textit{Randomisation} will have an impact over \textit{targeted} ads and can be adjusted according to user's needs i.e. `low-relevant vs. high-relevant interest-based ads', as is also evident in \cite{ullah2020protecting, tchen2014}; note that these latter set of techniques will also have impact over \textit{billing} since the advertisers' ads are shown to ``irrelevant'' users, hence, they (advertisers) pay for airtime that is used by non-targeted audiences. Similarly, an important parameter is the `trade-off between \textit{privacy} and \textit{targeted} ads', which can only be achieved using the \textit{Obfuscation} and the \textit{Randomisation} techniques. Furthermore, another parameter is to protect user privacy in terms of serving \textit{targeted ads} i.e. an `indirect privacy attack to expose user privacy', which cannot be exposed when \textit{Crypto-based} techniques are used since the delivered ads are also protected, as shown in \cite{ullah2017enabling}.

\subsection{The economic aspects of privacy}
Research works also investigate the notion of compensating users for their privacy loss, rather than imposing limits on the collection and use of personal information.

Ghosh and Roth \cite{ghosh2013selling} studied a market for private data, using \textit{differential privacy} as a measure of the privacy loss. The authors in \cite{riederer2011sale} introduce transactional privacy, which enables the users to sell (or lease) selected personal information via an auction system. On a related topic of content personalisation and \textit{in-browser} privacy, in RePriv \cite{fredrikson2011repriv} the authors propose a system that fits into the concept of a marketplace for private information. Their system enables controlling the level of shared (local) user profile information with the advertising networks, or, more broadly, with any online entity that aims to personalise content.

\section{Open Research Issues}\label{open-issues}

In this section, we present various future research directions that require further attention from the research community i.e. diffusion of user data in Real Time Bidding (RTB) scenarios and associated privacy risks, the complicated operations of advertising system, the user-driven private mobile advertising systems and its private \textit{billing} mechanism.

\subsection{Diffusion of user tracking data}
A recent shift in the online advertising has enabled by the advertising ecosystem to move from ad networks towards ad exchanges, where the advertisers bid on impressions being sold in RTB auctions. As a result, the A\&A companies closely collaborate for exchanging user data and facilitate bidding on ad impressions and clicks \cite{ bashir2016tracing, melicher2016not}. In addition, the RTB cause A\&A companies to perform additional tasks of working with publishers to help manage their relationship for ad exchange (in addition to user's tracking data) and to optimise the ad placement (i.e. \textit{targeted} ads) and bidding on advertiser's behalf. This has made the online advertising operations and the advertising ecosystems themselves extremely complex.

Hence, it is important for the A\&A companies to model (in order to accurately capture the relationship between publisher and A\&A companies) and evaluate the impact of RTB on the diffusion of user tracking (sensitive) data. This further requires assessing the advertising impact on the user's contexts and \textit{profiling} interests, which is extremely important for its applicability and scalability in the advertising scenarios. This will also help the A\&A companies and publisher to effectively predict the tracker domain and to estimate their advertising revenue. Furthermore, to ensure the privacy of user data since the data is collected and disseminated in a distributed fashion i.e. users affiliated to different \textit{analytics} and advertising platforms and shared their data across diverse publishers. This also necessitates a distributed platform for the efficient management and sharing of distributed data among various A\&A platforms and publishers. In particular, the RTB has demanded to develop efficient methods for distributed and private data management.

\subsection{Complex operations of advertising system}
The complexity of online advertising poses various challenges to user privacy, processing-intensive activities, interactions with various entities (such as CDN, \textit{analytics} servers, etc.) and their tracking capabilities. In order to reduce the complexity of the advertising systems, we envision few more areas of research: devising processing-sensitive frameworks, limiting the direction-redirection of requests among A\&A entities, unveil user data exchange processes within the ad platform, identifying new privacy threats and devising new protection mechanisms. Unveiling user data exchange will expose the extent to which the intermediate entities prone to adversarial attacks. Hence, it requires a better knowledge of adversary, which will contribute to develop protection mechanisms for various kinds of privacy threats, such as, interest-based attacks, direct privacy attacks. Note that this will further require comparative analysis of basic and new proposals for the trade-off achieved between privacy and computing overheads of processing user's ad retrieval requests/responses, communication bandwidth consumption and battery consumption.

\subsection{Private user-driven mobile advertising systems}
An enhanced user-driven private advertising platform is required where the user interest (vis-\`a-vis their privacy) and advertising system's business interests may vary, in addition, the assessment of user information as an inherent economic value will help to study the tradeoff between such values and user privacy within the advertising system. This will require the proposal for complex machine learning techniques to enhance ads \textit{targeting} (since previous works found that majority of received ads were not tailored to intended user profiles \cite{ullah2014characterising, nath2015madscope}, which will ultimately help advertising systems to increase their revenues and enhance user experience in receiving relevant ads. Likewise, introducing novel privacy preserving mechanisms, a very basic step would be to combine various proposals, as described in Section \ref{solutions}, which will introduce more robust and useful privacy solutions for various purposes: enhanced user \textit{targeting}, invasive tracking behaviors, better adapting privacy enhancing technologies, better adapt the changing economic aspects and \textit{ethics} in ads \textit{targeting}. Another research direction would be to extend the analysis of privacy protection mechanisms to other different players, such as, advertisers, ad exchange, publishers with the aim to analyse and evaluate privacy policies and protection mechanisms that are claimed by these parties. This would help various entities in the advertising system to identify the flaws and further improve their working environment.

Another research direction would be to create smarter privacy protection tools on the user side i.e. to create such tools as an essential component of mobile/browser-based platform within the advertising ecosystem. To develop such tools where users effectively enforce various protection strategies, it require various important parameters of usability, flexibility, scalability etc., to be considered to give users transparency and control over their private data.

Another research direction would be to extend the analysis of privacy protection mechanisms to other different players, such as, advertisers, ad exchange, publishers with the aim to analyse and evaluate privacy policies and protection mechanisms that are claimed by these parties. This would help various entities in the advertising system to identify the flaws and further improve their working environment.

\footnotetext{It \cite{dnt2009christopher} proposes a DNT field in the HTTP header that requests a web application to either disable the tracking (where it is automatically set) or cross-site the user tracking of an individual user.}

\subsection{Private billing mechanism}
Billing for both \textit{ad presentations} and \textit{clicks} is an important component of online advertising system. As discussed in Appendix \ref{private-billing}, a private \textit{billing} proposal is based on \textit{Threshold BLS signature}, \textit{Polynomial commitment}, and \textit{Zero knowledge proof} (ZKP), which are based on PIR mechanisms and \textit{Shamir secret sharing} scheme along with \textit{Byzantine robustness}. The applicability of this private \textit{billing} model can be verified in the online advertising system, which would require changes on both the user and ad system side. Furthermore, note that the this private \textit{billing} mechanism, implemented via \textit{polynomial commitment} and \textit{zero-knowledge proof}, is highly resource consuming process, henceforth, an alternative implementation with reduced processing time and query request size can be achieved via implementing together \textit{billing} with PIR using \textit{multi-secret sharing} scheme. In addition, to explore the effect of \textit{multi-secret sharing} scheme in multiple-server PIR and hence comparative analysis to choose between the two variations of \textit{single-secret} and \textit{multi-secret sharing} system implementations. \textit{Multi-secret sharing} scheme would help reduce the communication bandwidth and delays along with the processing time of query requests/responses

In addition, our \textit{billing} mechanism for \textit{ad presentations} and \textit{clicks} presented in \cite{ullah2017enabling}, also described in Section \ref{billing-ads}, is applicable only to single ad requests with no impact on privacy. However, the broader parameter values (simultaneously processing multiple ad requests) and the use of other PIR techniques, such as Hybrid-PIR \cite{devet2014best} and Heterogeneous-PIR \cite{mozaffariheterogeneous}, can be used to efficiently make use of processing time.

Furthermore, with the rise in popularity of Cryptocurrencies, many businesses and individuals have started investing in them, henceforth, the applicability of embedding the Cryptocurrency with the existing \textit{billing} methods needs an investigation and developing new frameworks for coexisting the \textit{billing} payments with the Cryptocurrency market. In addition, this would require techniques for purchasing, selling, and transferring Cryptocurrency among various parties i.e. ad systems, \textit{app} developers, publishers, advertisers, crypto-markets, and miners. A further analysis would require investigating the impact of such proposals on the current advertising business model with/without a significant effect.

An important research direction is to explore implementation of private advertising systems in Blockchain networks since there is limited Blockchain-based advertising systems e.g., \cite{gu2018secure, ullah2020privacy}. The \cite{ullah2020privacy} presents the design of a decentralised framework for \textit{targeted} ads that enables private delivery of ads to users whose behavioral profiles accurately match the presented ads, defined by the advertising systems. This framework provides: a private \textit{profiling} mechanism, privately requesting ads from the advertising system, the \textit{billing} mechanisms for ads monetisation, uploading ads to the cloud system, various types of transactions to enable advertising operations in Blockchain-based network, and \textit{access policy} over cloud system for accessing various resources (such as ads, mobile user profiles). However, its applicability in an actual environment is still questionable, in addition to, the coexistence of \textit{ads-billing} mechanism with Cryptocurrency.

\section{Conclusion} \label{conclusion}

Targeted/Online advertising has become ubiquitous on the internet, which has triggered the creation of new internet ecosystems whose intermediate components have access to billions of users and to their private data. The lack of transparency of online advertising, the A\&A companies and their operations have posed serious risks to user privacy. In this article, we break down the various instances of \textit{targeted} advertising, their advanced and intrusive tracking capabilities, the privacy risks from the information flow among various advertising platforms and ad/analytics companies, the \textit{profiling} process based on user's private data and the \textit{targeted} ads delivery process. Several solutions have been offered in the literature to help protect user privacy in such a complex ecosystem, henceforth, we provide a wide range of mechanisms that were classified based on the privacy mechanisms used, ad serving paradigm and the deployment scenarios (browser and mobile). Some of the solutions are very popular among internet users, such as blocking, however their blocking mechanism negatively impacts the advertising systems. On the other hand, majority of the proposals provide naive privacy that require a lot of efforts from the users; similarly, other solutions demand structural changes with the advertising ecosystems. We have found that it is very hard, based on various privacy preserving approaches, while demanding for devising novel approaches, to provide user privacy that could give users more control over their private data and to reduce the financial impact of new systems without significantly changing the advertising ecosystems and their operations.

\appendices
\section{Private Information Retrieval (PIR)}\label{pir-mechanisms}
PIR  \cite{chor1997computationally, kushilevitz1997replication, goldberg2007improving, Chor:1995:PIR:795662.796270, chor1997private, devet2014best} is a multiparty cryptographic protocol that allows users to retrieve an item from the database without revealing any information to the database server about the retrieved item(s). In one of our previous works \cite{ullah2017enabling}, our motivation for using PIR rather than other solutions, e.g., Oblivious Transfer \cite{chu2008efficient, naor1999oblivious}, is the lower communication and computation overheads of such schemes.

A user wishes to privately retrieve ${\beta ^{th}}$ record(s) from the database $D$. $D$ is structured as $r \times s$, where $r$ is the number of records, $s$ the size of each record; $s$ may be divided into words of size $w$. For \textit{multi-server} PIR, a scheme uses $l$ database servers and has a privacy level of $t$; $k$ is the number of servers that respond to the client's query, among those, there are $v$ \texttt{Byzantine} servers (i.e., malicious servers that respond incorrectly) and $h$ honest servers that send a correct response to the client's query. Following, we briefly discuss and compare various PIR schemes.

\subsection{Computational PIR (CPIR)}
The \textit{single-server} PIR schemes, such as CPIR \cite{aguilar2007lattice}, rely on the computational complexity (under the assumption that an adversary has limited resources) to ensure privacy against malicious adversaries. To privately retrieve the ${\beta ^{th}}$ record from $D$, a CPIR client creates a matrix ${M_\beta }$ by adding hard noise (based on large disturbance by replacing each diagonal term in ${M_\beta }$ by a random bit of $2^{40}$ words \cite{aguilar2007lattice}) to the desired record and soft noise (based on small disturbance) to all the other records. The client assumes that the server cannot distinguish between the matrices with hard and soft noises. The server multiplies the query matrix ${M_\beta }$ to the database ${D}$ that results in corresponding response ${R}$; the client removes the noise from ${R}$ to derive the requested record ${\beta ^{th}}$.

\subsection{Recursive CPIR (R-CPIR)}
The CPIR mechanism is further improved in terms of communication costs \cite{aguilar2007lattice} by recursively using the \textit{single-server} CPIR where the database is split into a set of virtual small record sets each considered as a virtual database. The query is hence calculated against part of the database during each recursion. The client recursively queries for the virtual records, each recursion results in a virtual database of smaller virtual records, until it determines a single (actual) record that is finally sent to the client.

\subsection{Information Theoretic PIR (IT-PIR)}
The \textit{multi-server} IT-PIR schemes \cite{goldberg2007improving, henry2011practical, beimel2004reducing, gertner1998random, devet2012optimally} rely on multiple servers to guarantee privacy against colluding adversaries (that have unbounded processing power) and additionally provide \textit{Byzantine robustness} against malicious servers.

To query a database for ${\beta ^{th}}$ record with protection against up to $t$ colluding servers, the client first creates a vector ${e_\beta }$, with `1' in the ${\beta ^{th}}$ position and `0' elsewhere. The client then generates $\left( {l,t} \right)$ \textit{Shamir secret shares} \sloppy $v_1, v_2, \cdots , v_l$ for ${e_\beta }$. The shares (one each) are subsequently distributed to the servers. Each server $i$ computes the response as \sloppy ${R_i} = {v_i} \cdot D$, this is sent back to the client. The client reconstructs the requested ${\beta ^{th}}$ record of the database from these responses. The use of of \textit{Shamir secret sharing} enables the recovery of the desired record from (only) $k \le l$ server responses \cite{goldberg2007improving}, where $k > t$ (and $t < l$).

\subsection{Hybrid-PIR (H-PIR)}
The \textit{multi-server} H-PIR scheme \cite{devet2014best} combines \textit{multi-server} IT-PIR \cite{goldberg2007improving} with the recursive nature of the \textit{single-server} CPIR \cite{aguilar2007lattice} to improve performance, by lowering the computation and communication costs\footnote{A complete implementation of CPIR, IT-PIR and H-PIR, \textit{Percy++} is present on \url{http://percy.sourceforge.net/}.}. Let these two schemes be respectively represented by $\tau $ for IT-PIR and the $\gamma $ for the recursive CPIR protocol. A client wants to retrieve ${\beta ^{th}}$ record then the client must determine the index of virtual records containing the desired records at each step of the recursion until the recursive depth $d$. The client creates an IT-PIR $\tau$-query for the first index and sends it to each server. It then creates CPIR $\gamma$-query during each of the recursive steps and sends it to all the servers. Similarly, on the server side at each recursive steps; the server splits the database into virtual records each containing actual records, uses the $\tau $ server computation algorithm, and finally uses the $\gamma $ CPIR server computation algorithm. The last recursive step results in the record ${R_i}$, that is sent back to the client.

\subsection{Comparison and applicability of various PIR techniques in ad systems}
Following comparative analysis, based on literature work, would help the selection of various PIR schemes and their applicability within an advertising system. We note that various performance metrics relate to the size of query along with the selection of a particular PIR scheme e.g., the CPIR takes longer processing delays and highest bandwidth consumption compared to both the IT-PIR and H-PIR schemes. This is due to the computations involved in query encoding and due to the servers performing \textit{matrix-by-matrix} computations instead of \textit{vector-by-matrix}, as is used by the IT-PIR and H-PIR schemes \cite{devet2014best}, although, the communication cost can be lowered down using the recursive version of the CPIR \cite{aguilar2007lattice}.

Furthermore, IT-PIR provides some other improvements, such as the \textit{robustness}, which is its ability to retrieve correct records even if some of the servers do not respond or reply with incorrect or malicious responses \cite{devet2012optimally}. It is further evident \cite{devet2014best} that both the \textit{single-server} CPIR and the \textit{multi-server} IT-PIR schemes, such as \cite{beimel2004reducing, goldberg2007improving, gertner1998random, henry2011practical}, respectively make the assumptions of computationally bounded and that particular thresholds of the servers are not colluding to discover the contents of a client's query. Alternatively, the H-PIR \cite{devet2014best}, provides improved performance by combining \textit{multi-server} IT-PIR with the recursive nature of \textit{single-server} CPIR schemes respectively to improve the computation and communication costs.

A recent implementation i.e., Heterogeneous PIR \cite{mozaffariheterogeneous}, enables \textit{multi-server} PIR protocols (implemented using multi-secret sharing algorithm, compatible with \textit{Percy++}\footnote{\url{http://percy.sourceforge.net/}} PIR library) over non-uniform servers (in a heterogeneous environment where servers are equipped with diverse resources e.g. computational capabilities) that impose different computation and communication overheads. This implementation makes it possible to run PIR over a range of different applications e.g. various resources (ad's contents such as, \texttt{JPEG}, \texttt{JavaScript} files) present on CDN in distributed environments. Furthermore, this implementation has tested and compared its performance with Goldberg's \cite{goldberg2007improving} implementation with different settings e.g., for different database sizes, numbers of queries and for various degrees of heterogeneity. This implementation achieves a trade-off between computation and communication overheads in heterogeneous server implementation by adjusting various parameters.

\section{Building blocks for enabling PIR and private billing}\label{private-billing}

This section introduces various building blocks for enabling PIR techniques i.e. \textit{Shamir secret sharing} and \textit{Byzantine robustness}. It further discusses various techniques that are used for private \textit{billing} i.e. \textit{Threshold BLS signature}, \textit{Polynomial commitment}, and \textit{Zero-knowledge proof} (ZKP).

\subsection{Shamir secret sharing}
The \textit{Shamir secret sharing} \cite{shamir1979share} scheme divides a \textit{secret} $\sigma $ into parts, giving each participant e.g. $l$ servers a unique part where some or all of the parts are needed in order to reconstruct the \textit{secret}. If the \textit{secret} is found incorrect then it can be handled through error-correcting codes, such as the one discussed in \cite{guruswami2008explicit}. Let the $\sigma $ be an element of some finite field $F$ then the \textit{Shamir scheme} works as follows: a client selects an $l$ distinct non-zero elements  ${\alpha _1},{\alpha _2}, \cdots ,{\alpha _l} \in F$ and selects $t$ elements ${a_1},{a_2}, \cdots ,{a_t}{ \in _R}F$  (the ${ \in _R}$ means uniformly at random). A polynomial $f\left( x \right) = \sigma  + {a_1}x + {a_2}{x^2} +  \cdots  + {a_t}{x^t}$ is constructed and gives the share $\left( {{\alpha _i},f\left( {{\alpha _i}} \right)} \right) \in F \times F$ to the server $i$ for $1 \le i \le l$. Now any $t + 1$ or more servers can use Lagrange interpolation \cite{devet2012optimally} to reconstruct the polynomial $f$ and, similarly, obtains $\sigma $ by evaluating $f\left( 0 \right)$.

\subsection{Byzantine robustness}
The problem of \textit{Byzantine} failure allows a server to continue its operation but it incorrectly responds. The \textit{Byzantine} failure may include corrupting of messages, forging messages, or sending conflicting messages through malice or errors. In order to ensure the responses' integrity in a \textit{single-server}, such as PIR-Tor \cite{mittal2011pir}, the server can provide a \textit{cryptographic signature} on each database's block. However, in a \textit{multi-server} PIR environment, the main aim of the \textit{Byzantine robustness} is to ensure that the protocol still functions correctly even if some of the servers fail to respond or provide incorrect or malicious responses. The client at the same time might also be interested in figuring out which servers have sent incorrect responses so that they can be avoided in the future.

The \textit{Byzantine robustness} for PIR was first considered by Beimel and Stahl \cite{beimel2003robust, beimel2007robust}; the scheme called the  $t$-private  $v$-\textit{Byzantine robust}  $k$-out-of-$l$ PIR. The authors take the  $l$-server information-theoretic PIR setting where $k$ of the servers respond, $v$ servers respond incorrectly, and the system can sustain up to $t$ colluding servers without revealing client's query among them. Furthermore, they suggest the \textit{unique decoding} where the protocol always outputs a correct unique block under the conditions $v \le t \le {k \mathord{\left/ {\vphantom {k 3}} \right. \kern-\nulldelimiterspace} 3}$.

The \cite{goldberg2007improving} uses the \textit{list decoding}, that is an alternative to unique decoding of error-correcting codes for large error rates, and demonstrates that the privacy level can be substantially increased up to $0 < t < k$ and the protocol can tolerate up to $k - \left\lfloor {\sqrt {kt} } \right\rfloor  - 1$ \textit{Byzantine} servers. Alternatively, the \textit{list decoding} can also be converted to \textit{unique decoding} \cite{micali2005optimal} at the cost of slightly increasing the database size \cite{devet2012optimally}.

Following schemes are the essential building blocks for enabling private \textit{billing} along with evaluating the PIR techniques for privately retrieving ads from the ad database.

\subsection{Threshold BLS signature}
The \textit{Boneh-Lynn-Shacham} (BLS) \cite{boneh2001short} is a `short' \textit{signature} verification scheme that allows a user to verify that the signer is authentic. The signer's private signing key is a random integer  $x \in {Z_q}$ and the corresponding public verification key is $\left( {\hat g, \hat g^{x}} \right)$ ($\hat g $ is a generator of ${\mathbb{G}_2}$). The procedure for \textit{signature} verification is as follows: Given the signing key $x$ and a message $m$, the \textit{signature} is computed via $\sigma  = {h_x}$ where $h = hash(m)$ is a cryptographic hash of $m$; the verification equation is $e(\sigma ,\hat g)\begin{array}{*{20}{c}} ?\\  =  \end{array}e(h, \hat g^{x})$, which results in true/false. To fit into scenario of multiple PIR servers; a  $\left( {k,l} \right)$-threshold variant of \textit{BLS signature} can be used where signing keys are the evaluations of a polynomial of degree $\left( k - l \right)$ and the master \textit{secret} is the constant term of this polynomial. Similarly, the reconstruction process can be done using Lagrange interpolation. The $\left( k - l \right)$ threshold \textit{BLS signature} partly provides the level of \textit{robustness} against the \textit{Byzantine} signers since the \textit{signature} share can be verified independently using the signer's public verification key share.

\subsection{Polynomial commitment}
A \textit{polynomial commitment} \cite{kate2010constant} scheme allows committers to formulate a constant-sized \textit{commitments} to polynomials that s(he) can commit so that it can be used by a verifier to confirm the stated evaluations of the committed polynomial \cite{kate2010polynomial}, without revealing any additional information about the committed value(s). An example of the \textit{Polynomial commitment} constructions in \cite{kate2010constant} provides unconditional hiding if a \textit{commitment} is opened to at most $t - 1$ evaluations (i.e. $t - 1$ servers for a degree-$t$ polynomial) and provides computational hiding under the discrete $log (DL)$ if polynomial \textit{commitment} is opened to at least $t$ evaluations. As presented in \cite{kate2010constant}, \textit{commitment} to a polynomial  $f\left( x \right) = {a_t}{x^t} +  \cdots  + {a_1}z + {a_0}$ has the form  ${{\cal C}_f} = {\left( {{g^{{\alpha ^t}}}} \right)^{{a_t}}} \cdots {\left( {{g^\alpha }} \right)^{{a_1}}}{g^{{a_0}}} = {g^{f\left( \alpha  \right)}}$ where $\alpha $ is \textit{secret},  $g \in {\mathbb{G}_1}$ is a generator whose discrete logarithm with respect to $g$ is unknown, including all the bases are part of the \textit{commitment} scheme's \textit{public key}. The verifier, on the other side, can confirm that the claimed evaluations is true by checking if $Ver\left( {{{\cal C}_f},r,f\left( r \right),w} \right) = \left[ {e\left( {{{\cal C}_f},{\hat g}} \right)\mathop  = \limits^? e\left( {w,{{{\hat g^{\alpha }}} \mathord{\left/
 {\vphantom {{{\hat g^{\alpha }}} {{\hat g^{r}}}}} \right.
 \kern-\nulldelimiterspace} {{\hat g^{r}}}}} \right).e{{\left( {g,{\hat g}} \right)}^{f\left( r \right)}}} \right]$ is \texttt{true}, here the \textit{commitment} $w$ is called the \textit{witness}; detailed discussion can be found in \cite{kate2010constant}.

\subsection{Zero-knowledge proof (ZKP)}
The \textit{zero knowledge proof} is an interactive protocol between the \textit{prover} and the \textit{verifier} that allows the \textit{prover} to prove to the \textit{verifier} that it holds a given statement without revealing any other information. There are several \textit{ZKPs}, such as range proof to prove that a committed value is non-negative \cite{boudot2000efficient}, the proof of knowledge of a committed value \cite{schnorr1990efficient}, knowledge proof of a discrete log representation of a number \cite{brands2000rethinking}, and proof that a \textit{commitment} opens to multiple \textit{commitments} \cite{camenisch1999proving}. Besides, there are several batch proof techniques, such as \cite{bellare1998fast, bellare1998batch} to achieve verification of a basic operation like modular exponentiation in some groups, which significantly reduces the computation time.

\section{k-anonymity}\label{k-anonymity}
\textit{k-anonymity} was introduced in \cite{samarati1998protecting, samarati2001protecting} and its enforcement through generalization and suppression was suggested in \cite{sweeney2002k}. \textit{k-anonymity} examines the re-identification attack, which aims to release private version of the data (i.e. structured data e.g. data holders of bank or hospital etc.) that cannot be re-identified while the data still remains useful. Let \(RT\left( {{A_1}, \ldots ,{A_n}} \right)\) be a set of structured data organised in rows and columns, a population of entities $U$, with a finite set of attributes of \(RT\)as \(\left( {{A_1}, \ldots ,{A_n}} \right)\) with at least one attribute identified as `\textit{key attribute}' that can be considered as \textit{quasi-identifier}\footnote{Variable values or combinations of variable values within a dataset that are not structural uniques but might be empirically unique and therefore in principle uniquely identify a population unit. \url{https://stats.oecd.org/glossary/detail.asp?ID=6961}}\footnote{Quasi-identifiers are pieces of information that are not of themselves unique identifiers, but are sufficiently well correlated with an entity that they can be combined with other quasi-identifiers to create a unique identifier. \url{https://en.wikipedia.org/wiki/Quasi-identifier}}. A \textit{quasi-identifier} of $RT$, represented as \({Q_{RT}}\), is a set of attributes \(\left( {{A_1}, \ldots ,{A_j}} \right) \subseteq \left( {{A_1}, \ldots ,{A_n}} \right)\), where \(\exists {p_i} \subset U\) such that \({f_g}\left( {{f_c}\left( {{p_i}} \right)\left[ {{Q_{RT}}} \right]} \right) = {p_i}\); \({f_c}:U \to RT\) and \({f_g}:RT \to U'\), \(U \subseteq U'\).

\textit{k-anonymity} for $RT$ is achieved if each sequence of values in \(RT\left[ {{Q_{RT}}} \right]\) appears with at least $k$ occurrences i.e. \({Q_{RT}} = \left( {{A_1}, \ldots ,{A_j}} \right)\) be the \textit{quasi-identifier} associated with $RT$, where \({A_1}, \ldots ,{A_j} \subseteq {A_1}, \ldots ,{A_n}\) and $RT$ satisfy \textit{k-anonymity}. Subsequently, each sequence of values in \(RT\left[ {{A_x}} \right]\) appears with at least $k$ occurrences in \(RT\left[ {{Q_{RT}}} \right]\) for \(x = i, \ldots ,j\). The $RT$ satisfies the \textit{k-anonymity} is released. The combination of any set of attributes of the released data $RT$ and external sources on which \({{Q_{PT}}}\) ($PT$ is the private table) is based, cannot be linked that eventually guarantees the privacy of released data. A detailed example is given in \cite{samarati2001protecting}.

\bibliographystyle{ieeetr}
\bibliography{references}

\begin{IEEEbiography}[{\includegraphics[width=1in,height=1.25in,clip,keepaspectratio]{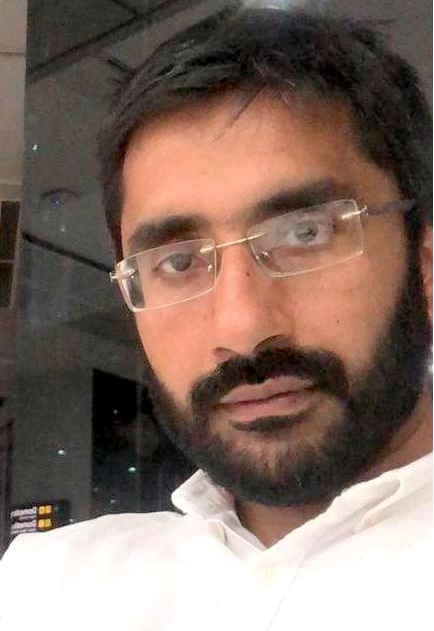}}]{Imdad Ullah} (Member, IEEE) has received his Ph.D. in Computer Science and Engineering from The University of New South Wales (UNSW) Sydney, Australia. He is currently an assistant professor with the College of Computer Engineering and Sciences, PSAU, Saudi Arabia. He has served in various positions of Researcher at UNSW, Research scholar at National ICT Australia (NICTA)/Data61 CSIRO Australia, NUST Islamabad Pakistan and SEEMOO TU Darmstadt Germany, and Research Collaborator at SLAC National Accelerator Laboratory Stanford University USA. He has research and development experience in privacy preserving systems including private advertising and crypto-based billing systems. His primary research interest include privacy enhancing technologies; he also has interest in Internet of Things, Blockchain, network modeling and design, network measurements, and trusted networking.
\end{IEEEbiography}

\begin{IEEEbiography}[{\includegraphics[width=1in,height=1.25in,clip,keepaspectratio]{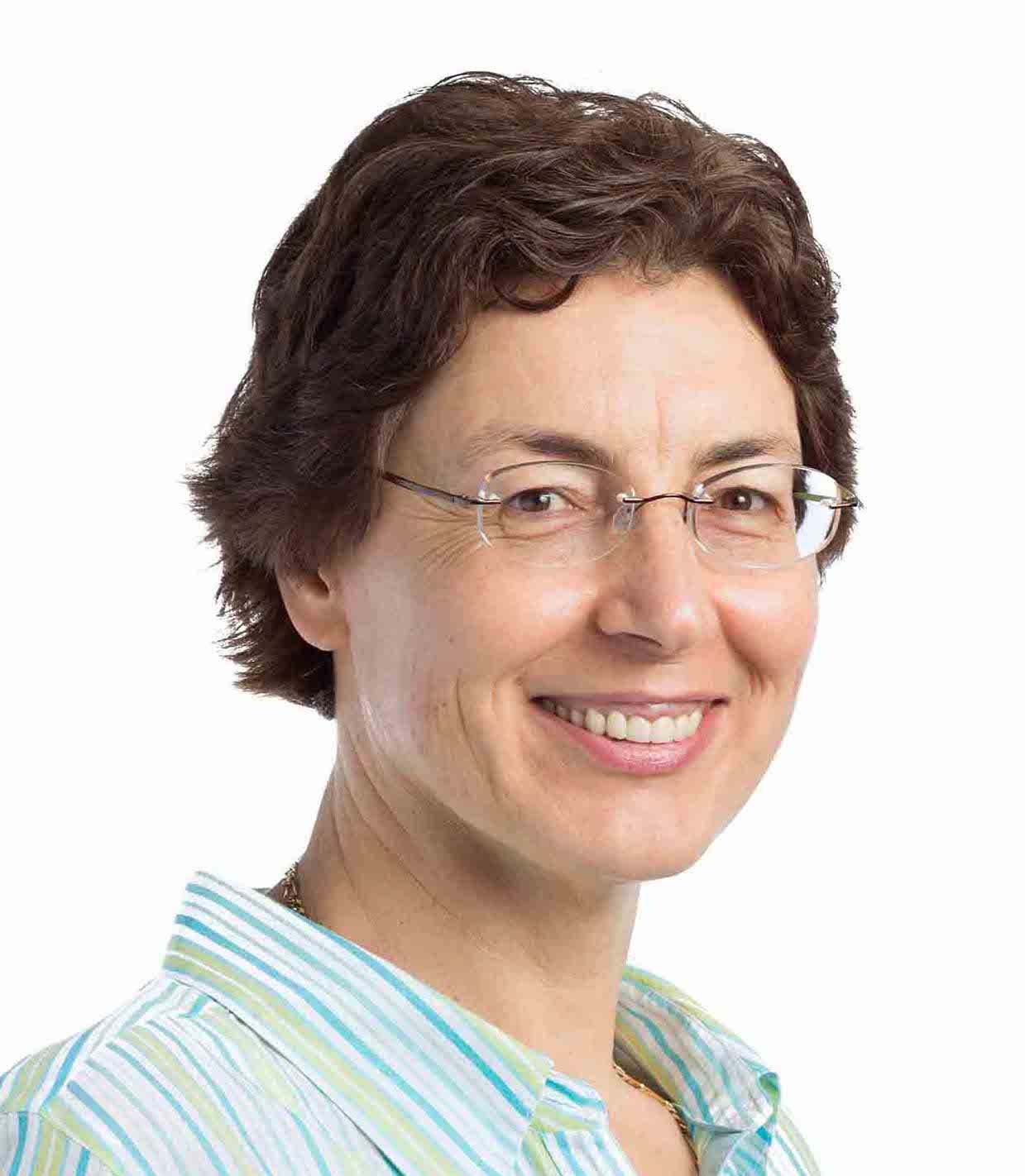}}]{Roksana Boreli} has received her Ph.D in Communications from University of Technology, Sydney, Australia.  She has over 20 years of experience in communications and networking research and in engineering development, in large telecommunications companies (Telstra Australia, Xantic, NL) and research organisations. Roksana has served in various positions of Engineering manager, Technology strategist, Research leader of the Privacy area of Networks research group in National ICT Australia (NICTA)/CSIRO Data61 and CTO in a NICTA spinoff 7-ip. Her primary research focus is on the privacy enhancing technologies; she also maintains an interest in mobile and wireless communications.
\end{IEEEbiography}

\begin{IEEEbiography}[{\includegraphics[width=1in,height=1.25in,clip,keepaspectratio]{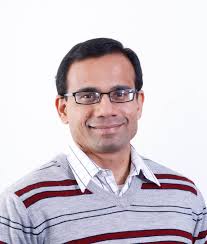}}]{Salil S. Kanhere} (Senior Member, IEEE) received the M.S. and Ph.D. degrees from Drexel University, Philadelphia. He is currently a Professor of Computer Science and Engineering with UNSW Sydney, Australia. His research interests include the Internet of Things, cyberphysical systems, blockchain, pervasive computing, cybersecurity, and applied machine learning. He is a Senior Member of the ACM, an Humboldt Research Fellow, and an ACM Distinguished Speaker. He serves as the Editor in Chief of the Ad Hoc Networks journal and as an Associate Editor of the IEEE Transactions On Network and Service Management, Computer Communications, and Pervasive andMobile Computing. He has served on the organising committee of several IEEE/ACM international conferences. He has co-authored a book titled \textit{Blockchain for Cyberphysical Systems}.
\end{IEEEbiography}




\end{document}